\documentclass[12pt]{article}
\usepackage{amsfonts}
\usepackage{amsmath}
\usepackage{amssymb}
\usepackage{array}
\usepackage{bigints}
\usepackage{booktabs}
\usepackage[nosort]{cite}
\usepackage{color}
\usepackage{dsfont}
\usepackage{float}
\usepackage{framed}
\usepackage{graphicx}
\usepackage{indentfirst}
\usepackage{mathrsfs}
\usepackage{multirow}
\usepackage{setspace}
\usepackage{subdepth}
\usepackage{subfig}
\usepackage{titlesec}
\usepackage[dotinlabels]{titletoc}
\usepackage{wrapfig}
\usepackage[all]{xy}
\usepackage{young}
\usepackage[vcentermath]{youngtab}

\usepackage{hyperref}
\hypersetup{colorlinks=true}
\hypersetup{linkcolor=black}
\hypersetup{citecolor=black}
\hypersetup{urlcolor=black}

\numberwithin{equation}{section}

\usepackage[left=2.5cm,right=2.5cm,top=2.5cm,bottom=3cm]{geometry}
\titleformat{\section}{\normalfont\bfseries}{\thesection.}{4pt}{}
\titlespacing{\section}{0pt}{20pt}{6pt}

\titleformat{\subsection}{\normalfont\itshape}{\thesubsection.}{4pt}{}
\titlespacing{\subsection}{0pt}{15pt}{6pt}

\titleformat{\subsubsection}{\normalfont}{\thesubsubsection.}{4pt}{}
\titlespacing{\subsubsection}{0pt}{15pt}{6pt}

\def\ie{\begin{equation}\begin{aligned}}
\def\fe{\end{aligned}\end{equation}}

\newcommand\VRule[1][\arrayrulewidth]{\vrule width #1}

\def\tilde{\widetilde}
\def\t{\tilde}
\def\hat{\widehat}

\def\half{{1 \over 2}}
\def\d{\partial}

\def\1{{\mathds 1}}

\DeclareMathOperator{\tr}{tr}
\DeclareMathOperator{\Tr}{\mathrm{Tr}}

\newcommand{\Z}{{\mathbb Z}}

\newcommand{\R}{{\mathbb R}}

\def\SP{{\mathscr P}}

\def\CI{{\mathcal I}}

\def\CL{{\mathcal L}}
\def\CM{{\mathcal M}}
\def\CN{{\mathcal N}}

\def\CT{{\mathcal T}}

\DeclareFontShape{OT1}{cmr}{mx}{n}%
    {<->cmr10}{}
\newcommand{\mytitlefont}{\fontseries{mx}\selectfont}
\DeclareMathAlphabet{\titlemath}{OT1}{cmr}{mx}{n}

\begin{document}


\begin{titlepage}

\begin{center}

~\\[2cm]

{\fontsize{24pt}{0pt} \mytitlefont  2-Group Global Symmetries and Anomalies\\[5pt]  in Six-Dimensional Quantum Field Theories}

~\\[0.1cm]

Clay C\'{o}rdova,$^1$ Thomas T.~Dumitrescu,$^2$ and Kenneth Intriligator\,$^3$

~\\[0.1cm]

$^1$\,{\it Kadanoff Center for Theoretical Physics \& Enrico Fermi Institute, University of Chicago}\\[4pt]

$^2$\,{\it Mani L.\,Bhaumik Institute for Theoretical Physics, Department of Physics and Astronomy,}\\[-2pt]
       {\it University of California, Los Angeles, CA 90095, USA}\\[4pt]

$^3$\,{\it Department of Physics, University of California, San Diego}

~\\[10pt]

\end{center}

\noindent We examine six-dimensional quantum field theories through the lens of higher-form global symmetries. Every Yang-Mills gauge theory in six dimensions, with field strength $f^{(2)}$, naturally gives rise to a continuous 1-form global symmetry associated with the 2-form instanton current $J^{(2)} \sim * \Tr \left( f^{(2)} \wedge f^{(2)}\right)$. We show that suitable mixed anomalies involving the gauge field $f^{(2)}$ and ordinary 0-form global symmetries, such as flavor or Poincar\'e symmetries, lead to continuous 2-group global symmetries, which allow two flavor currents or two stress tensors to fuse into the 2-form current $J^{(2)}$. We discuss several features of 2-group symmetry in six dimensions, many of which parallel the four-dimensional case. The majority of six-dimensional supersymmetric conformal field theories (SCFTs) and little string theories have infrared phases with non-abelian gauge fields. We show that the mixed anomalies leading to 2-group symmetries can be present in little string theories, but that they are necessarily absent in SCFTs. This allows us to establish a previously conjectured algorithm for computing the 't Hooft anomalies of most SCFTs from the spectrum of weakly-coupled massless particles on the tensor branch of these theories. We then apply this understanding to prove that the $a$-type Weyl anomaly of all SCFTs with a tensor branch must be positive, $a > 0$. 

\vfill

\begin{flushleft}
September 2020
\end{flushleft}

\end{titlepage}


\tableofcontents

\section{Introduction and Summary}

The detailed study of string constructions in which gravity can be decoupled in a controlled way strongly indicates the existence of many non-gravitational, interacting, and UV complete theories in $d = 6$ spacetime dimensions. All known 6d theories in this class preserve $\CN=(1,0)$, $\CN=(2,0)$, or $\CN=(1,1)$ supersymmetry (SUSY), as well as Poincar\'e symmetry. This class of theories can be further divided into two subclasses:

\begin{itemize}
\item[1.)] Superconformal field theories (SCFTs) are renormalization group (RG) fixed points, without an intrinsic scale, that are believed to obey the rules of local quantum field theory (QFT) \cite{Seiberg:1996qx}. (See~\cite{Gaiotto:2014kfa} for a modern  discussion of some aspects related to this, and \cite{Heckman:2018jxk} for a recent survey of string constructions for 6d SCFTs.) In particular, they possess a well-defined, local stress tensor $T_{\mu\nu}(x)$. Together with unitarity and superconformal symmetry, this restricts the allowed SUSY algebras to $\CN=(1,0)$ and $\CN=(2,0)$ (see~\cite{Cordova:2016emh} and references therein), as realized in all known examples.

With the exception of free $\CN=(1,0)$ tensor or hyper multiplets, all known 6d SCFTs are strongly coupled and do not appear to have a Lagrangian description. However, all known interacting 6d SCFTs possess a tensor branch -- a moduli space of vacua along which conformal symmetry is spontaneously broken, parametrized by the vacuum expectation values (vevs) of scalars residing in $\CN=(1,0)$ tensor multiplets. At low energies, these tensor multiplets are weakly coupled, and they may be accompanied by other weakly-coupled matter fields in $\CN=(1,0)$ hyper or vector multiplets. Note that a free $\CN=(1,0)$ vector multiplet is not an SCFT, because a free Maxwell field is not conformally invariant in six dimensions. Going onto the moduli space of vacua by activating vevs is the only way to initiate a supersymmetric RG flow out of a 6d SCFT, since such theories do not possess SUSY-preserving relevant or marginal operators~\cite{Louis:2015mka,Cordova:2016xhm}.

\item[2.)] Little string theories (LSTs) are UV complete, interacting theories without gravity that are not local quantum field theories. They can be obtained by taking $M_\text{Planck} \rightarrow \infty$  while keeping $M_\text{string}$ finite in suitable string or brane constructions~\cite{Seiberg:1997zk} (see \cite{Aharony:1999ks} for a review). A typical signature for the non-locality of LSTs is the fact that they enjoy T-duality \cite{Seiberg:1997zk} and an asymptotic Hagedorn density of states \cite{Maldacena:1997cg}, both of which are stringy features associated with energies $E \gtrsim M_\text{string}$. At low energies, $E \ll M_\text{string}$, LSTs flow to conventional local QFTs perturbed by irrelevant local operators, i.e.~they are described by a standard low-energy effective field theory. The low-energy QFT may either be an SCFT, in which case the super-Poincar\'e symmetry must be $\CN=(1,0)$ or $\CN=(2,0)$, or it may include IR free non-conformal sectors, such as SUSY Yang-Mills theories constructed using only $\CN = (1,0)$ vector and hyper (but not tensor) multiplets. In the latter case the super-Poincar\'e symmetry can also be $\CN=(1,1)$ rather than $\CN=(1,0)$ or $\CN=(2,0)$. 

The non-locality of LSTs obstructs the definition of well-behaved local operators, such as a local stress tensor $T_{\mu\nu}(x)$. For this reason, the basic observables in LSTs are momentum-space correlators such as $\langle T_{\mu\nu}(p) T_{\rho\sigma}(-p)\rangle$. These correlators grow exponentially at large momenta (roughly due to the Hagedorn density of states) and this renders their Fourier transform to position space ill defined \cite{Peet:1998wn,Minwalla:1999xi}. However, they appear to be  sufficiently well behaved to allow for the definition of approximately local macroscopic observables at distances much longer than $M^{-1}_\text{string}$ \cite{Kapustin:1999ci}.

\end{itemize}

\noindent The 6d theories described above have been studied through a combination of string and field theoretic techniques. Since these theories are typically difficult to analyze head on, symmetries (as well as associated anomalies) have played an important role in elucidating their properties.

 In this paper we examine the 6d SUSY theories above, along with other (generally non-UV-complete and non-SUSY) low-energy effective QFTs in 6d, through the lens of higher-form global symmetries \cite{Kapustin:2014gua, Gaiotto:2014kfa}. The reason these make a natural appearance here is that every 6d gauge theory, with Yang-Mills field strength $f^{(2)}$, naturally gives rise to a continuous\footnote{~Aspects of discrete 1-form symmetries in 6d SUSY theories were recently discussed in~\cite{Morrison:2020ool,Apruzzi:2020zot,Bhardwaj:2020phs}.} 1-form global symmetry $U(1)^{(1)}$ associated with the 2-form instanton current,\footnote{~We essentially follow the conventions of \cite{Cordova:2018cvg}. In particular, we work in Euclidean signature, and we use a superscript $X^{(n)}$ to indicate that $X$ is an $n$-form. Moreover, abelian gauge fields are taken to be hermitian, while non-abelian ones are taken to be valued in the anti-hermitian Lie algebra of the gauge group.}
\begin{equation}
J^{(2)} \sim * \Tr \left( f^{(2)} \wedge f^{(2)}\right)~,
\end{equation}
where $*$ is the Hodge star (see section \ref{ssec:oneformsym} below). As was shown in \cite{Cordova:2018cvg}, mixed anomalies involving the gauge field $f^{(2)}$ and ordinary 0-form symmetries, such as flavor or Poincar\'e symmetries, make it possible for the operator product expansion (OPE) of flavor currents or stress tensors to contain 2-form currents such as $J^{(2)}$. This current algebra structure, which mixes conserved currents associated with 0-form and 1-form global symmetries, was termed continuous 2-group global symmetry in \cite{Cordova:2018cvg}.\footnote{~Just as there are discrete higher-form global symmetries \cite{Kapustin:2014gua, Gaiotto:2014kfa}, there are also discrete 2-group global symmetries. They were explicitly identified in \cite{Kapustin:2013uxa} and further studied in \cite{Kapustin:2014zva, Thorngren:2015gtw, Bhardwaj:2016clt, Tachikawa:2017gyf, Benini:2018reh,Hsin:2020nts}. There are useful parallels between continuous and discrete 2-group global symmetries, e.g.~both can be interpreted as variants of the Green-Schwarz mechanism that mix ordinary and higher-form background gauge fields (see section \ref{ssec:2gpreview} below). However, the interpretation in terms of current algebra only applies to the continuous case.} The analysis in \cite{Cordova:2018cvg} focused on $d = 4$, where continuous 1-form global symmetries, and hence continuous 2-group global symmetries, only arise in a somewhat special subset of theories -- most notably in abelian gauge theories. Here we generalize this discussion to $d = 6$, where continuous 2-group global symmetries can also occur in the substantially richer setting of non-abelian gauge theories. Such non-abelian gauge theories are known to arise in many IR phases of 6d SCFTs and LSTs. As we will see, a detailed understanding of 2-group global symmetries in these theories will allow us to clarify some of their puzzling properties that have emerged in recent years.

\subsection{Continuous 1-Form Global Symmetries in 6d Gauge Theories}\label{ssec:oneformsym}

As was already mentioned above, we are primarily interested in continuous 1-form global symmetries in $d = 6$, but we begin by reviewing a few general facts about such symmetries that apply to all $d$ (see see~\cite{Kapustin:2014gua,Gaiotto:2014kfa} and references therein). 

A continuous 1-form global symmetry $U(1)_B^{(1)}$ is associated with a conserved 2-form current $J^{(2)}_B$. It's Hodge dual $*J^{(2)}$ is a closed $(d-2)$-form, $d * J^{(2)} = 0$. In components,
\begin{equation}\label{eq:jbcons}
J_B^{\mu\nu} = J_B^{[\mu\nu]}~, \qquad \d_\mu J_B^{\mu\nu} = 0~.
\end{equation}
Topological (and hence conserved) charge operators can be defined by integrating $*J^{(2)}_B$ over a $(d-2)$-surface $\Sigma_{d-2}$,\footnote{~The factor of $i$ in \eqref{eq:qbdef}, as well as in~\eqref{instcurr} below, is due to our Euclidean conventions.}
\begin{equation}\label{eq:qbdef}
Q_B(\Sigma_{d-2}) =  -i \int_{\Sigma_{d-2}} * J_B^{(2)}~.
\end{equation}
Note that $J_B^{(2)}$ naturally has dimension $d-2$, so that the integrated charges $Q_B(\Sigma_{d-2})$ are dimensionless. This is necessarily the case in CFTs, where unitarity bounds imply that any operator $J_B^{(2)}$ satisfying \eqref{eq:jbcons} must have scaling dimension $\Delta(J_B^{(2)}) = d-2$ (see for instance \cite{Cordova:2016emh}). If the charges are quantized, $Q_B \in \Z$, then the 1-form symmetry is compact (i.e.~it is $U(1)^{(1)}_B$ rather than $\R^{(1)}_B$), and we will only consider the compact case here. 

The operators or defects that carry $Q_B$ charge must be line defects, which can be linked by a closed $\Sigma_{d-2}$. If the $Q_B$ symmetry is not spontaneously broken, the action of such line operators on the vacuum creates a dynamical string that is charged under $Q_B$. In this case $\Sigma_{d-2}$ are the $d-2$ spatial dimensions transverse to the string worldsheet. In supersymmetric theories, $Q_B$ can appear in the SUSY algebra, which makes it possible for strings that carry $Q_B$ charge to be BPS. In this case $J_B^{(2)}$ resides in the stress-tensor supermultiplet of the theory (see \cite{Dumitrescu:2011iu} and references therein). 

It is useful to couple the current $J_B^{(2)}$ (as well as other currents or operators of interest) to suitable classical sources. In general, conserved currents are sourced by background gauge field, and the appropriate source for $J_B^{(2)}$ is a 2-form background gauge field $B^{(2)}$,\footnote{~As in \cite{Cordova:2018cvg}, we denote background gauge fields such as $B^{(2)}$ with uppercase letters, while dynamical gauge fields such as $f^{(2)}$ are denoted using lowercase letters.}
 \begin{equation}\label{BwedgeJ}
S \quad \supset \quad \int B^{(2)} \wedge * J^{(2)}_B~.
\end{equation}
The fact that $J_B^{(2)}$ is a conserved current amounts to invariance under background gauge transformations $B^{(2)} \rightarrow B^{(2)}+d\Lambda ^{(1)}_B$, where $\Lambda^{(1)}_B$ is the corresponding 1-form gauge parameter.\footnote{~More precisely, $\Lambda_B^{(1)}$ is a 1-form gauge field, with quantized periods ${1 \over 2 \pi} \int_{\Sigma_2} d \Lambda_B^{(1)}  \in \Z$.}

Perhaps the simplest continuous higher-form symmetries arise as ``magnetic'' symmetries in standard Yang-Mills gauge theories, possibly with electrically charged matter, which does not spoil the magnetic symmetries. As long as there are no dynamical magnetic charges, the gauge-covariant Bianchi identity implies that the Yang-Mills field strength $f^{(2)}$ is a covariantly closed 2-form. Every rank-$k$ Casimir of the gauge algebra $\frak g$ gives rise to a gauge-invariant Chern class; schematically,
\begin{equation}\label{eq:ckdef}
c_k(f^{(2)}) \sim \Tr \left( (f^{(2)})^{\wedge k}\right)~.
\end{equation}
Here $k$ is a positive integer, which is bounded in terms of the spacetime dimension, $2k \leq d$. Then $c_k(f^{(2)})$ is a closed $(2k)$-form of dimension $2k$.  It's Hodge dual thus defines a conserved $(d-2k)$-form current,
\begin{equation}\label{eq:ckjdef}
J^{(d-2k)} = i * c_k(f^{(2)})~, \qquad d * J^{(d-2k)} = 0~.
\end{equation}
Since the $c_k(f^{(2)})$ are normalized so that their periods are quantized, the same is true of the charges $Q_B$ defined in \eqref{eq:qbdef}. Thus the current $J^{(d-2k)}$ gives rise to a compact $(d-2k-1)$-form symmetry $U(1)^{(d-2k -1)}$. 

An even simpler variant occurs in dynamical $U(1)$ gauge theory with Maxwell field strength $f^{(2)}$. In the absence of dynamical magnetic charges, the Bianchi identity implies that the first Chern class $c_1(f^{(2)}) = {1 \over 2 \pi} {f^{(2)} }$ is a closed, gauge-invariant 2-form with quantized periods. Consequently, $J^{(d-2)} = i * c_1(f^{(2)})$ is a conserved $(d-2)$-form current that gives rise to a $U(1)^{(d-3)}$ symmetry. In the $d = 4$ examples studied in \cite{Cordova:2018cvg} this was a $U(1)^{(1)}$ symmetry; in the $d= 6$ examples analyzed below it is a $U(1)^{(3)}$ symmetry. Moreover, $U(1)$ gauge theory in 6d also has a $U(1)^{(1)}$ symmetry associated with $J^{(2)} = {i \over 4 \pi^2} * \left(f^{(2)}  \wedge f^{(2)} \right)$.\footnote{~Note that $Q(\CM_4) = {1 \over 4 \pi^2} \int_{\CM_4} f^{(2)} \wedge f^{(2)} \in \Z$ for any closed, oriented 4-manifold $\CM_4$, while $Q(\CM_4) \in 2\Z$ is even if $\CM_4$ is also spin. For a general 4-cycle $\Sigma_4 \subset \CM_6$ inside a spin 6-manifold $\CM_6$ we do not expect a stronger quantization condition than $Q(\Sigma_4) \in \Z$. This explains our choice of normalization for $J^{(2)}$ in abelian gauge theories.}
 
In this paper, we are primarily interested in the $U(1)_B^{(1)}$ global symmetry that naturally arises in IR free 6d Yang-Mills gauge theories, with gauge algebra $\frak g$, via the second Chern class $c_2(f^{(2)})$. This corresponds to setting $d = 6$ and $k = 2$ in \eqref{eq:ckdef} and \eqref{eq:ckjdef}. More precisely,
\begin{equation}\label{instcurr}
J^{(2)}_B = i*c_2(f^{(2)})~, \qquad c_2(f^{(2)}) =  \frac{1}{8 \pi^2}\Tr \Big(f^{(2)} \wedge f^{(2)}\Big)~.
\end{equation}
Here $\Tr = {1 \over 2 h^\vee_\frak{g}} \Tr_\text{adj}$, where $h^\vee_\frak{g}$ is the dual Coxeter number of  $\frak g$, while $\Tr_\text{adj}$ is the trace in the adjoint representation. The normalization is chosen so that a minimal $\frak g$-instanton on $S^4$ satisfies $\int_{S^4} c_2(f^{(2)}) = 1$. In other words, $c_2(f^{(2)})$ is the conventionally normalized instanton density. Therefore the $U(1)_B^{(1)}$ symmetry in 6d is associated with instantons of the dynamical Yang-Mills gauge field. (For this reason, we will occasionally refer to it as the instanton 1-form symmetry.) The dynamical charged string excitations are the familiar instanton string solitons in 6d gauge theory. Perhaps less familiar are the charged line defects: they are instanton disorder lines, which are defined by the requirement that any small $S^4$ linking the line should carry a certain fixed instanton number. Note that the instanton symmetry $U(1)_B^{(1)}$ is not spontaneously broken in IR free 6d gauge theories. To see this, note that the  current $J_B^{(2)} \sim \Tr(f^{(2)} \wedge f^{(2)})$ creates at least two gluons from the vacuum, rather than a single Nambu-Goldstone (NG) particle. By contrast, the $U(1)^{(d-3)}$ magnetic symmetry associated with an abelian gauge field (see above) is spontaneously broken in the Coulomb phase, since the associated current $J^{(d-2)} \sim *f^{(2)}$ creates a single NG particle -- the photon -- from the vacuum. 

When a 6d theory with instanton 1-form symmetry is circle-compactified to 5d, it gives rise to an ordinary 0-form symmetry associated with instanton particle solitons and instanton disorder local operators. The latter are the 5d analogue of 3d monopole operators, which are also associated with a particular 0-form symmetry.  Monopole and instanton 0-form symmetries play an important role in the dynamics of 3d and 5d gauge theories, and they have been particularly well studied in the supersymmetric context. Even though these symmetries are naturally abelian, they can combine with other 0-form symmetries into a larger, non-abelian flavor symmetry \cite{Intriligator:1996ex,Seiberg:1996bd}. Somewhat reminicently, we will see below that even though 6d instanton 1-form symmetries are (like all 1-form symmetries~\cite{Gaiotto:2014kfa}) abelian, they can combine with ordinary 0-form symmetries into a larger 2-group symmetry.

\subsection{Review of Continuous 2-Group Global Symmetries}\label{ssec:2gpreview}

As was already mentioned above, it is possible for the conserved currents associated with continuous 0-form symmetries to fuse into a 2-form current $J_B^{(2)}$ associated with a continuous 1-form symmetry $U(1)_B^{(1)}$. The resulting structure is a continuous 2-group global symmetry. Here we will review those aspects of continuous 2-group symmetries that we will need below, following the discussion in \cite{Cordova:2018cvg}, which contains many additional details and examples. For now, we do not fix the spacetime dimension $d$.

Consider a continuous 0-form flavor symmetry $G^{(0)}$, which (for now) we take to be a compact simple group. The corresponding conserved current is $j_G^{(1)}$; it is valued in the Lie algebra of $G^{(0)}$. Another 0-form symmetry, which is present in every relativistic QFT, is Poincar\'e symmetry ${\mathscr{P}}^{(0)}$. The associated conserved current is the stress tensor $T_{\mu\nu}$. In many examples, these 0-form symmetries are unaffected by the presence of the 1-form symmetry $U(1)_B^{(1)}$, which we emphasize by writing the global symmetry as an ordinary product,
\begin{equation}\label{eq:prodsym}
G^{(0)} \times {\mathscr{P}}^{(0)} \times U(1)_B^{(1)}~.
\end{equation}
In this case the currents $j_G^{(1)}$ and $T_{\mu\nu}$ satisfy conventional conservation laws,
\begin{equation}\label{eq:ordjgtcons}
d * j_G^{(1)} = 0~, \qquad T_{\mu\nu} = T_{(\mu\nu)}~, \qquad \d^\mu T_{\mu\nu} = 0~.
\end{equation}
This in turn implies that they can be coupled to background fields in a standard way: a background 1-form gauge field $A_G^{(1)}$ for $j_G^{(1)}$ (both are valued in the Lie algebra of $G^{(0)}$), and a background Riemannian geometry for $T_{\mu\nu}$. We choose to describe the latter by a background vielbein $e^{(1)a} = e_\mu^a dx^\mu$, so that the Riemannian background metric is $g_{\mu\nu} = \delta_{ab} e^a_\mu e^b_\nu$. The conservation equations \eqref{eq:ordjgtcons} for the currents are reflected in the following transformation rules for the background gauge fields,
\begin{equation}\label{eq:aegt}
A^{(1)}_G \rightarrow  A^{(1)}_G+d_A \lambda _G^{(0)}~, \qquad e^{(1)a} \rightarrow e^{(1)a}-{\theta ^{(0)a}}_b e^{(1)b}+\CL _\xi e^{(1)a}~.
\end{equation}
Here $d_A = d + [ A^{(1)}_G, \, \cdot \, ]$ is the $G^{(0)}$-covariant exterior derivative and $\lambda_G^{(0)}$ parametrizes an infinitesimal $G^{(0)}$ background gauge transformation, while ${\theta ^{(0)a}}_b$ parametrizes an infinitesimal local Lorentz transformation and $\xi = \xi^\mu \d_\mu$ an infinitesimal diffeomorphism for the background geometry. As in section \ref{ssec:oneformsym}, the 2-form current $J_B^{(2)}$ satisfies the conservation equation
\begin{equation}\label{eq:jbconsbis}
d * J_B^{(2)} = 0~,
\end{equation}
and couples to a background 2-form gauge field $B^{(2)}$ subject to background gauge transformations parametrized by the 1-form gauge parameter $\Lambda_B^{(1)}$,
\begin{equation}\label{eq:bgt}
B^{(2)} \rightarrow B^{(2)} + d \Lambda^{(1)}_B~.
\end{equation}

A 2-group global symmetry arises when the currents, and hence the background fields, for 0-form and 1-form symmetries mix in a specific way. We emphasize this by writing the global symmetry as a twisted product,
\begin{equation}\label{eq:twistedsym}
\left(G^{(0)}\times \SP^{(0)}\right) \times _{\hat \kappa _G, \hat \kappa _\SP} U(1)^{(1)}~.
\end{equation}  
Here $\hat \kappa_G, \hat \kappa_\SP$ are constants (in fact integers, see below) that characterize the 2-group symmetry. We refer to them as 2-group structure constants, in analogy with Lie algebra structure constants. If there are additional 0-form or 1-form symmetries to consider, they generally lead to additional 2-group structure constants. 

The twisted natured of the 2-group symmetry \eqref{eq:twistedsym} manifests in several closely related ways. Perhaps the most easily recognizable one involves the background gauge fields: while the background gauge transformations for $A_G^{(1)}$ and $e^{(1)a}$ are unchanged, and of the form~\eqref{eq:aegt}, the background gauge transformations for $B^{(2)}$ are modified: in addition to the transformations parametrized by the 1-form gauge parameter $\Lambda_B^{(1)}$ in \eqref{eq:bgt}, $B^{(2)}$ also shifts under the $G^{(0)}$ background gauge transformations parametrized by $\lambda_G^{(0)}$ and under local Lorentz transformations parametrized by ${\theta^{(0)a}}_b$,\footnote{~As in the familiar Green-Schwarz mechanism, there is some freedom in the choice of transformation rule for $B^{(2)}$, which is related to local counterterms. For instance, we choose to shift $B^{(2)}$ under local Lorentz transformations, but not under diffeomorphisms. By adding a local counterterm, we can instead have $B^{(2)}$ shift under diffeomorphisms, but not under local Lorentz transformations. See \cite{Cordova:2018cvg} for additional details.} 
\begin{equation}\label{eq:Bshiftsintro}
B^{(2)} \rightarrow B^{(2)} +d\Lambda _B^{(1)}  +\frac{\hat \kappa _G}{4\pi}\Tr\left( \lambda _G^{(0)}dA_G^{(1)}\right)+\frac{\hat \kappa _\SP}{16\pi} \tr \left(\theta ^{(0)}d\omega ^{(1)}\right)~.
\end{equation}
Here $\hat \kappa_G, \hat \kappa_\SP$ are the 2-group structure constants in \eqref{eq:twistedsym}, ${\omega^{(1)a}}_b$ is the spin connection associated with $e^{(1)a}$, and $\tr$ denotes a trace over local Lorentz indices. Note that $dA^{(1)}$ and $d \omega^{(1)}$ involve the ordinary exterior derivative, i.e. these expressions do not correspond to the field strength of $A_G^{(1)}$ or the Riemann curvature of $\omega^{(1)}$.  The shift in \eqref{eq:Bshiftsintro} takes exactly the same form as in the Green-Schwarz (GS) mechanism.\footnote{~From a mathematical point of view the GS transformation law \eqref{eq:Bshiftsintro} can be viewed as defining a 2-connection associated with the 2-group symmetry. See \cite{baez2004higher,Baez:2004in,Sati:2008eg,Schreiber:2008} for further details.}

A key difference is that the familiar GS mechanism applies to dynamical gauge fields (i.e.~all fields involved are dynamical), while all gauge fields in \eqref{eq:Bshiftsintro} are background gauge fields.\footnote{~There is also an intermediate variant of the GS mechanism, where $B^{(2)}$ is replaced by a dynamical 2-form gauge field $b^{(2)}$, while the gauge fields that appear in the GS shift of $b^{(2)}$ may be either dynamical or background fields (see section \ref{ssec:gsintro} below).} This leads to important conceptual and practical differences. 

A familiar consequence of the modified transformation rule \eqref{eq:Bshiftsintro} is that $d B^{(2)}$ is no longer invariant under background $G^{(0)}$ gauge transformations or local Lorentz transformations. Instead, it is standard to define a modified field strength $H^{(3)}$ via
\begin{equation}\label{Hshifts}
{1 \over 2 \pi} H^{(3)} = {1 \over 2 \pi} d B^{(2)} -{ \hat \kappa _G \over 8 \pi^2} \, \text{CS}(A^{(1)}_G)-{\hat \kappa _\SP \over 32 \pi^2} \,  \text{CS} (\omega^{(1)})~.
\end{equation}
Here $\text{CS}(A^{(1)}_G)$ and $\text{CS} (\omega^{(1)})$ are Chern-Simons terms for the background gauge and gravity fields, 
\begin{equation}\label{eq:csdef}
\text{CS}(A^{(1)}_G) = \Tr \left( A^{(1)}_G \wedge d A^{(1)}_G + {2 \over 3} (A^{(1)}_G)^{\wedge 3}\right)~, \; \text{CS}(\omega^{(1)}) = \tr \left (\omega^{(1)} \wedge d \omega^{(1)} + {2 \over 3} (\omega^{(1)})^{\wedge 3}\right)~,
\end{equation}
which transform as follows under background $G^{(0)}$ gauge transformations, and local Lorentz transformations, respectively,
\begin{equation}\label{eq:csvar}
\text{CS}(A^{(1)}_G) \rightarrow \text{CS}(A^{(1)}_G) + d \Tr \left(\lambda_G^{(1)} d A_G^{(1)}\right)~, \qquad \text{CS}(\omega^{(1)}) \rightarrow \text{CS}(\omega^{(1)}) + d \tr \left(\theta^{(0)} d \omega^{(1)}\right)~.
\end{equation}
It follows that $H^{(3)}$ is completely invariant under all background gauge and gravity transformations, but it is no longer closed. Instead, 
\begin{equation} \label{eq:dhnotzero}
{1 \over 2 \pi} d H^{(3)} = -\hat \kappa _G \, c_2(F^{(2)}_G) + {\hat \kappa _\SP \over 4}  \, p_1(T)~,
\end{equation}
with $c_2(F_G^{(2)}) = {1 \over 8 \pi^2} \Tr \left(F_G^{(2)} \wedge F_G^{(2)}\right)$ the second Chern class of the $G^{(0)}$ bundle (see \eqref{instcurr}), defined in terms of the field strength $F_G^{(2)} = d_A A_G^{(1)}$, and $p_1(T) = - {1 \over 8 \pi^2} \tr \left(R^{(2)} \wedge R^{(2)}\right)$ the first Pontryagin density of the tangent bundle, defined in terms of the Riemann curvature ${R^{(2)a}}_b = d {\omega^{(1)a }}_b + {\omega^{(1)a}}_c \wedge {\omega^{(1)c}}_b$. As explained in \cite{Cordova:2018cvg}, the consistency of equations such as \eqref{eq:Bshiftsintro}, \eqref{Hshifts}, and \eqref{eq:dhnotzero} require the 2-group structure constants $\hat \kappa_G, \hat \kappa_\SP$ to be quantized, 
\begin{equation}\label{kappaquant}
\hat \kappa _G\in \Z~, \qquad \hat \kappa _\SP \in \Z~.
\end{equation}

An important difference between 2-group global symmetry and the familiar GS mechanism is that 2-group symmetry implies modified conservation laws for the currents $j_A^{(1)}$ and $T_{\mu\nu}$ associated with the 0-form global symmetries $G^{(0)}$ and $\SP^{(0)}$. These modified conservation laws, which can be read off from the GS shifts of $B^{(2)}$ in \eqref{eq:Bshiftsintro},\footnote{~\label{currnorm}This is true up to a prefactor that depends on how we normalize the coupling of background fields to currents. We use the conventions of \cite{AlvarezGaume:1984dr,Cordova:2018cvg}: the variation of the (effective) action under a $G^{(0)}$ gauge transformation parametrized by $\lambda_G^{(0)}$ is $S \rightarrow S -2 \int \Tr \left(\lambda_G^{(0)}  \, d_A * j_G^{(1)}\right) + (\text{GS  shift of } B^{(2)})$; its variation under a local Lorentz transformation parametrized by $\theta^{(0)}_{ab} = \theta^{(0)}_{[ab]}$ is $S \rightarrow S - \int  \theta^{(0) ab} \, *T_{[ab]} + (\text{GS  shift of } B^{(2)})$.} involve the 2-form current operator $J_B^{(2)}$, whose conservation law is unchanged, as well as the background gauge fields for $G^{(0)}$ and $\SP^{(0)}$, 
\begin{equation}\label{modcons}
d_A * j_G^{(1)}=  \frac{\hat\kappa _G}{8\pi} \, d A_G^{(1)} \wedge *J_B^{(2)}~, \qquad * T_{[ab]}=  -\frac{\hat \kappa _\SP}{16\pi}  \, d \omega^{(1)}_{[ab]}\wedge * J_B^{(2)}~.
\end{equation}
These equations show that the standard conservation law for $j_A^{(1)}$ and the symmetry of $T_{\mu\nu}$ are violated in the presence of background fields. Moreover, these violations involve the current $J_B^{(2)}$, which is a non-trivial operator. On the other hand, these violations disappear if we set the background fields to zero. The non-conservation equations in \eqref{modcons} are therefore somewhat in between conventional 't Hooft anomalies (which involve only background fields on the right-hand side) and Adler-Bell-Jackiw (ABJ) anomalies (which involve only operators): the fact that the currents are unmodified in the absence of background fields means that the corresponding symmetries are not broken (just as for 't Hooft anomalies, and unlike ABJ anomalies, since the latter explicitly break symmetries). As was explained in \cite{Cordova:2018cvg}, the modified conservation laws in the presence of background fields lead to a deformation of the symmetry at the level of current algebra. This leads to an alternative, but fully equivalent, presentation of 2-group symmetry. 

To arrive at this presentation, we differentiate the modified conservation equations \eqref{modcons} with respect to the background gauge fields a second time, and then set the background fields to zero. Doing this for $j_G^{(1)}$ leads to 
\begin{equation}\label{djjope}
\d_\mu j_G^{\mu a} (x) j^{\nu b}_G(y) + f^{abc} \delta^{(d)}(x-y) j^{\nu c}_G(x) = {\hat \kappa_G \over 8 \pi} \delta^{ab} {\d \over \d x^\lambda} \delta^{(d)}(x-y) \; J_B^{\lambda\nu}(x)~.
\end{equation}
Here $a, b$ are $G^{(0)}$ adjoint indices, and $f^{abc}$ are the structure constants of its Lie algebra. If $\hat \kappa_G = 0$, so that the right-hand side of \eqref{djjope} vanishes, this OPE captures the fact that the current $j_G^{\nu b}(y)$ transforms in the adjoint representation of $G^{(0)}$. Integrating \eqref{djjope} to remove the derivative on $j^{\mu a}_G(x)$, we recover the standard fact that the $j^{(1)a}_G(x) j^{(1)b}_G(y)$ OPE contains a term $\sim f^{abc} j_G^{(1)c}$ at separated points, which encodes the Lie algebra of $G^{(0)}$. 

If the 2-group structure constant $\hat \kappa_G \neq 0$, the right-hand side of \eqref{djjope} implies that $\d_\mu j_G^{\mu a} (x) j^{\nu b}_G(y)$ contains the 2-form current operator $J_B^{(2)}$. This implies that the $j_G^{(1)a }(x) j_G^{(1)b}(y)$ OPE contains a term $\sim \hat\kappa_G \delta^{ab} J_B^{(2)}$ at separated points, which encodes the 2-group structure constant $\hat \kappa_G$. At this level, the 2-group current algebra is reminiscent of the non-abelian $G^{(0)}$ structure among the $j_G^{(1)}$ currents discussed above. An important difference is that the left-hand side of \eqref{djjope} contains a $\delta$-function, while the right-hand side contains the derivative of a $\delta$-function. This reflects the fact that $j_G^{(1)}$ is itself charged under $G^{(0)}$, while $J_B^{(2)}$ is not. By inserting \eqref{djjope} into correlation functions we can derive the Ward identities of the 2-group global symmetry \cite{Cordova:2018cvg}. 

We can repeat the preceding discussion for the stress-tensor equation in \eqref{modcons}. This leads to a formula similar to (but more complicated than) \eqref{djjope} for $T_{[\mu\nu]}(x) T_{\rho\sigma}(y)$, which can be analyzed along similar lines. For instance, it implies that the $T_{\mu\nu}(x) T_{\rho\sigma}(y)$ OPE contains a term $\sim \hat \kappa_\SP J_B^{(2)}$ at separated points.

\subsection{Basic Examples: 6d Gauge Theories with Mixed Gauge-Global Anomalies}\label{ssec:mixanom2gpintro}

Starting now, and for the remainder of this paper, we will only discuss theories in $d = 6$ spacetime dimensions. Adapting the discussion in \cite{Cordova:2018cvg} to six dimensions, we outline a simple, general mechanism through which continuous 2-group global symmetries can arise in 6d gauge theories with suitable anomalies. A more detailed discussion can be found in section \ref{ssec:moreImixed}.

Consider a dynamical Yang-Mills theory in six dimensions, with gauge group $g^{(0)}$ and field strength $f^{(2)}_g$, that couples to other degrees of freedom -- either to matter charged under $g^{(0)}$ (this could be weakly coupled scalars or fermions charged under $g^{(0)}$, or even strongly-coupled charged sectors) or to dynamical 2-form gauge fields (e.g.~via GS couplings). Assume that the charged matter also enjoys a 0-form flavor symmetry $G^{(0)}$, with background field strength $F_G^{(2)}$. To simplify this overview, we assume that $g^{(0)}$ and $G^{(0)}$ are compact simple Lie groups.\footnote{~For the most part, the global form of gauge and flavor groups will not play a prominent role below, i.e.~most of our results only depend on the Lie algebras of $g^{(0)}$ and $G^{(0)}$.} 

Via the standard descent procedure (see section \ref{ssec:moreImixed}), perturbative anomalies involving $g^{(0)}$ and $G^{(0)}$ are conveniently summarized in terms of an anomaly 8-form polynomial $\CI^{(8)}$, which is constructed from the Chern classes $c_k(f_g^{(2)})$ and $c_k(F_G^{(2)})$ for the dynamical and background gauge fields. (We will soon add background gravity fields as well.) It is useful to separate $\CI^{(8)}$ into three terms,
\begin{equation}\label{eq:i8def}
\CI^{(8)} = \CI^{(8)}_\text{gauge}[f^{(2)}_g] + \CI^{(8)}_\text{global}[F^{(2)}_G] + \CI^{(8)} _\text{mixed}[f^{(2)}_g,F^{(2)}_G]~.
\end{equation}
We discuss each term in turn:
\begin{itemize}
\item[1.)] The first term $\CI^{(8)}_\text{gauge}$ only depends on the dynamical gauge field~$f^{(2)}_g$, i.e.~it encodes pure gauge anomalies. These must necessarily vanish for consistency of the gauge theory, so that $\CI^{(8)}_\text{gauge} = 0$.

\item[2.)]  The second term $\CI^{(8)}_\text{global}$ only depends on the background gauge field $F^{(2)}_G$. Ordinarily, this term would encode 't Hooft anomalies of the global symmetry $G^{(0)}$, i.e.~interesting, scheme-independent observables of the theory that necessarily remain constant along RG flows. As was shown in \cite{Cordova:2018cvg} (see also section \ref{ssec:2gphooft}), not all terms in $\CI^{(8)}_\text{global}$ lead to meaningful 't Hooft anomalies in the presence of 2-group symmetry. 

\item[3.)] The third term $\CI^{(8)} _\text{mixed}$ depends on the dynamical gauge field $f^{(2)}_g$ and the background gauge field $F^{(2)}_G$. If $g^{(0)}$ and $G^{(0)}$ are both non-abelian, as we are currently assuming, the only possible term of this kind takes the form
\begin{equation}\label{eq:Imixedex}
\CI^{(8)} _\text{mixed}[f^{(2)}_g,F^{(2)}_G] = k_{g^2 G^2} \, c_2(f^{(2)}_g) \, c_2(F^{(2)}_G)~.
\end{equation}
The anomaly coefficient $k_{g^2 G^2}$ encodes a possible mixed anomaly between the dynamical $g^{(0)}$ and the background $G^{(0)}$ gauge symmetry. In a theory where $f^{(2)}_g$ and $F^{(2)}_G$ only couple to free fermions, $k_{g^2 G^2}$ can be computed from a box diagram with two external $g^{(0)}$ gluons, two external $G^{(0)}$ currents $j_G^{(1)}$, and fermions running in the loop. On general grounds, the anomaly coefficient must be quantized,\footnote{~\label{fn:csquantintro}This can be seen by demanding that the 7d Chern-Simons terms $S_\text{CS} = 2 \pi i \int_{\CM_7} \CI^{(7)}$, with $d \CI^{(7)} = \CI^{(8)}$, be properly quantized; equivalently, that $\int_{\CM_8} \CI^{(8)} \in \Z$ for 8-manifolds $\CM_8$ supporting $g^{(0)}$ and $G^{(0)}$ gauge fields (see section \ref{ssec:quantcond}). }
\begin{equation}\label{eq:kgGquant}
k_{g^2 G^2} \in \Z~.
\end{equation}

\end{itemize}

We will now argue that a genuinely six-dimensional theory with non-zero $k_{g^2 G^2} \neq 0$ necessarily has 2-group global symmetry (see section \ref{ssec:moreImixed} for additional details). Typically, theories with mixed anomalies such as \eqref{eq:Imixedex} can be regularized in such a way to preserve either one of the symmetries $g^{(0)}$, $G^{(0)}$ -- but not both. The difference between two such regularization schemes is accounted for by a 6d local counterterm constructed using the gauge fields $a_g^{(1)}$ and $A_G^{(1)}$. Since $g^{(0)}$ is a gauge symmetry, it must necessarily be anomaly free, even in the presence of $G^{(0)}$ background fields.  This leads to the following two possibilities:
\begin{itemize}
\item[(a)] The theory under consideration is genuinely six-dimensional, i.e.~the only dynamical fields propagate in 6d. In this case we must choose the aforementioned counterterm such that the mixed anomaly $\CI^{(8)} _\text{mixed}$ is completely pushed into the global $G^{(0)}$ symmetry. As we explain below, this leads to a 6d theory with 2-group global symmetry.

\item[(b)] The 6d theory under consideration lives at the boundary of a dynamical 7d bulk theory, or more generally  on a 6d defect embedded in a higher-dimensional bulk theory. In either case, the theory is not actually six-dimensional. Here it is not necessary to tune the 6d counterterm to cancel the $g^{(0)}$ anomaly arising from $\CI^{(8)} _\text{mixed}$ in the presence of $G^{(0)}$ background fields. Instead, this anomaly can be cancelled by anomaly inflow from the dynamical higher-dimensional bulk fields. For instance, we could extend the dynamical $g^{(0)}$ and background $G^{(0)}$ gauge fields into a 7d bulk. A certain 7d Chern-Simons term (obtained from $\CI^{(8)} _\text{mixed}$ via descent) can then cancel the mixed anomaly via inflow. In this case the mixed anomaly does not lead to 2-group global symmetry on the 6d boundary. 
\end{itemize}

\noindent In this paper we will not consider option (b)~above, i.e.~we take the view that all theories we study are genuinely six-dimensional. However, even if a theory is genuinely six-dimensional, with 2-group global symmetry in the deep IR, this symmetry may be emergent, i.e.~it may not be a symmetry of the full, microscopic 6d theory.\footnote{~There are interesting general constraints on emergent 2-group symmetries that are not present for conventional product symmetries \cite{Cordova:2018cvg}, e.g.~consider a 2-group $U(1)_G^{(0)} \times_{\hat \kappa_G} U(1)_B^{(1)}$. Because $B^{(2)}$ shifts under $A_G^{(1)}$ gauge transformations, but not vice versa, it is not possible for the $U(1)_G^{(0)}$ symmetry to emerge unless the $U(1)_B^{(1)}$ symmetry is already present. Therefore the energy scales $E_G, E_B$ at which these symmetries emerge must satisfy an inequality of the form $E_G \lesssim E_B$. See \cite{ClayBrennToApp} for an application of a similar inequality to axion models.} 

Once the mixed anomaly in \eqref{eq:Imixedex} has been regularized to preserve the dynamical $g^{(0)}$ gauge symmetry, the conservation law for the global $G^{(0)}$ current $j_G^{(1)}$ in the presence of a $G^{(0)}$ background field takes the following form (see section \ref{ssec:moreImixed}),
\begin{equation}\label{eq:GGggnonconsintro}
d_A * j_G^{(1)} = - {i k_{g^2 G^2} \over 8 \pi} dA_G^{(1)} \wedge c_2(f_g^{(2)})~.
\end{equation}
We identify $i c_2(f_g^{(2)}) = *J_B^{(2)}$, where $J_B^{(2)}$ is the 2-form current of the $U(1)_B^{(1)}$ instanton 1-form symmetry defined in~\eqref{instcurr}, which is conserved due to the Bianchi identity, $d*J_B^{(2)} = 0$. Then \eqref{eq:GGggnonconsintro} takes exactly the form of the first 2-group non-conservation equation in \eqref{eq:GGggnonconsintro}. The 2-group structure constant $\hat \kappa_G$ given in terms of the mixed anomaly coefficient,
\begin{equation}\label{eq:hkgiskgg}
\hat \kappa_G = -k_{g^2 G^2}~,
\end{equation}
which is compatible with the quantization conditions \eqref{kappaquant} and \eqref{eq:kgGquant}. Thus the mixed anomaly in \eqref{eq:Imixedex} fuses the global $G^{(0)}$ and $U(1)^{(1)}_B$ symmetries into the 2-group global symmetry $G^{(0)} \times_{\hat \kappa_G} U(1)_B^{(1)}$, with $\hat \kappa_G$ given by \eqref{eq:hkgiskgg}. As explained around \eqref{eq:Bshiftsintro}, this requires the background gauge field $B^{(2)}$ that sources $J_B^{(2)}$ to undergo a GS shift under $G^{(0)}$ background gauge transformations. In section \ref{ssec:moreImixed} we show how this GS shift of $B^{(2)}$ is simply inferred from the mixed anomaly. 

Let us briefly mention several variants of the preceding discussion that will also make an appearance below (some of these are spelled out in section  \ref{ssec:moreImixed}, and others can be worked out along similar lines):
\begin{itemize}
\item Analogously to \eqref{eq:Imixedex}, we can also consider a mixed anomaly between the dynamical $g^{(0)}$ gauge symmetry and Poincar\'e symmetry $\SP^{(0)}$,
\begin{equation}\label{Imixedeg}
\CI^{(8)}_\text{mixed} \, \supset  \, \frac{k _{g^2\SP^2}}{24}  \, c_2(f^{(2)}_g )\,  p_1(T)~.
\end{equation}
Here $p_1(T)$ is the first Pontryagin class of the tangent bundle, defined below \eqref{eq:dhnotzero}, and the normalization is such that the anomaly coefficient $k_{g^2 \SP^2}$ is always an integer,
\begin{equation}\label{eq:kgPquant}
k_{g^2 \SP^2} \in \Z~.
\end{equation}
The presence of such a mixed anomaly leads to the 2-group symmetry $\SP^{(0)} \times_{\hat \kappa_\SP} U(1)_B^{(1)}$, with 2-group structure constant\begin{equation}\label{eq:khpiskgp}
\hat \kappa_\SP = {\kappa_{g^2 \SP^2} \over 6}~,
\end{equation}
along with the corresponding gravitational GS shift for $B^{(2)}$ in \eqref{eq:Bshiftsintro} and the second 2-group non-conservation equation in \eqref{modcons}.  Note that \eqref{eq:khpiskgp} is not obviously compatible with the quantization conditions \eqref{kappaquant} and \eqref{eq:kgPquant}. This apparent discrepancy is explained in section \ref{ssec:quantcond}. 

\item We can consider variants of the mixed anomaly \eqref{eq:Imixedex} where either the dynamical $g^{(0)}$ gauge symmetry or the global $G^{(0)}$ symmetry is a $U(1)$ symmetry, in which case we must replace $c_2(f^{(2)}_g) \rightarrow {1 \over 4 \pi^2}  \left(f^{(2)}_g  \wedge f^{(2)}_g \right)$, and similarly for $F_G^{(2)}$, in the formulas above. (See section \ref{ssec:quantcond} for an abelian example.) 

Since sections \ref{sec:LSTs} and \ref{sec:scfts} are dedicated to 6d SUSY gauge theories, it is worth pointing out that there are no $\CN=(1,0)$ $U(1)$ gauge theories with charged hyper multiplets that do not also contain dynamical gravity. The reason is that such charged hypers necessarily lead to a non-vanishing, reducible $U(1)$ gauge anomaly, whose sign is such that the anomaly cannot be canceled using a GS mechanism involving chiral 2-form gauge fields residing in dynamical $\CN=(1,0)$ tensor multiplets. The only $\CN=(1,0)$ multiplet containing an anti-chiral 2-form gauge field that can cancel the anomaly is the supergravity multiplet, which must therefore be dynamical. 

\item Consider a dynamical $g^{(0)} = U(1)$ gauge theory with a mixed anomaly $\CI^{(8)} = \CI^{(6)}_\text{global} \wedge c_1(f_g^{(2)})$, where $\CI^{(6)}_\text{global}$ is an anomaly 6-form constructed only from background $G^{(0)}$ or gravity fields. Such an anomaly gives rise to a 4-group global symmetry, in which the $U(1)^{(3)}$ current $J^{(4)} = i * c_1(f_g^{(2)})$ discussed below \eqref{eq:ckjdef} mixes with the 0-form symmetries. This case can be analyzed along similar lines as the 6d and 4d 2-group cases discussed below and in \cite{Cordova:2018cvg}, respectively, but we will not do so explicitly.
\end{itemize}

\subsection{UV Complete Examples: 6d Little String Theories}\label{ssec:introlst}

The preceding discussion shows that 6d gauge theories with mixed anomalies such as \eqref{eq:Imixedex} and \eqref{Imixedeg} have continuous 2-group global symmetries. However, such theories are generically not UV complete. In section \ref{sec:LSTs} we present SUSY examples of precisely such gauge theories that have UV completions as 6d LSTs, demonstrating that 6d theories with continuous 2-group global symmetry can be UV completed. It is possible that these 2-group symmetries are emergent, i.e.~that they are accidental symmetries of the gauge theories in the deep IR, but not exact symmetries of the full LSTs.\footnote{~4d examples with emergent 2-group symmetries were presented in \cite{Cordova:2018cvg}.} A more interesting possibility is that the 2-group symmetries uncovered in the IR do extend to the full LSTs. See section \ref{sec:LSTs} for some additional comments, as well as the upcoming work~\cite{delzohmToApp}.

Rather than analyzing large classes of LSTs we focus on two examples studied in \cite{Seiberg:1997zk}:

\begin{itemize}
\item By taking the little-string decoupling limit for a stack of $N$ NS 5-branes in  type IIB string theory, we obtain a 6d LST with $\CN=(1,1)$ SUSY. At low energies this theory flows to  maximally supersymmetric Yang-Mills (MSYM) theory in 6d with gauge group $U(N)$. Decoupling the center-of-mass mode we obtain an $SU(N)$ gauge theory, with gauge algebra $A_{N-1}$. Other constructions also lead to $\CN=(1,1)$ LSTs flowing to 6d MSYM theories with gauge algebras $D_k$ and $E_k$ (see for instance \cite{Aharony:1999ks} and references therein).  

In addition to the $\CN=(1,1)$ super-Poincar\'e symmetry, all of these theories have an $SU(2)^{(0)}_L \times SU(2)^{(0)}_R$ symmetry under which the supercharges transform in the spin $(\half, \half)$ representation. We will show that these symmetries have mixed anomalies with the gauge symmetry, of the form \eqref{eq:Imixedex}, and hence they combine with the $U(1)_B^{(1)}$ instanton 1-form symmetry into a non-trivial 2-group.

\item By taking the little-string decoupling limit for the theory describing $N$ small, coincident $SO(32)$ instantons in the heterotic string, we find a 6d LST with $\CN=(1,0)$ SUSY. At low energies it flows to an $Sp(N)$ gauge theory with certain hyper multiplet matter~\cite{Witten:1995gx}.

In addition to the $\CN=(1,0)$ super-Poincar\'e symmetry, these theories have an $SU(2)^{(0)}_R$ symmetry under which the supercharges transform in the spin-$\half$ representation, as well as a $G^{(0)} = SU(2)_L^{(0)}\times SO(32)^{(0)}$ flavor symmetry that commutes with the supercharges. We will show that both $SU(2)_R^{(0)}$ and $G^{(0)}$ have mixed anomalies of the form \eqref{eq:Imixedex}, and that there is also a mixed gauge-gravity anomaly of the form \eqref{Imixedeg}. Thus the $SU(2)_R^{(0)} \times G^{(0)}$ symmetries, as well as the Poincar\'e symmetry $\SP^{(0)}$, are fused into a larger 2-group symmetry together with the $U(1)_B^{(1)}$ instanton 1-form symmetry.

\end{itemize}

\subsection{Absence of Continuous 1-Form and 2-Group Global Symmetries in 6d SCFTs}\label{ssec:no1fmscftint}

Unlike the LSTs discussed in section \ref{ssec:introlst} above, 6d SCFTs do not support continuous 2-group global symmetries. This is because they do not even admit continuous 1-form symmetries, which are necessary ingredients for continuous 2-group symmetries. 

As explained in section \ref{ssec:oneformsym}, a continuous 1-form global symmetry is associated with a 2-form current $J_B^{\mu\nu}$ satisfying the conservation law $\d_\mu J_B^{\mu\nu} = 0$ as an operator equation (i.e.~at separated points). In a unitary CFT, a non-trivial 2-form current must therefore be the conformal primary of a short, unitary representation of the conformal algebra,\footnote{\label{fn:trivcurr}~A non-trivial current, whose charge integrals do not automatically vanish, cannot be the total derivative of another well-defined local operator. In a CFT this means it cannot be a conformal descendant and must therefore be a conformal primary.} where the shortening condition corresponds to the conservation law $\d_\mu J_B^{\mu\nu} = 0$. The representation theory of the 6d conformal algebra does indeed allow for precisely such a short representation (i.e.~the existence of conserved 2-form currents are not in principle ruled out in 6d CFTs) as long as the operator $J_B^{\mu\nu}$ has operator dimension $\Delta = 4$, which is consistent with the discussion below \eqref{eq:qbdef}. 
 
However, in unitary SCFTs the local operators must reside in unitary representations of the appropriate superconformal algebras. A systematic approach to the problem of enumerating conformal primary operators residing in such superconformal representations was presented in \cite{Cordova:2016xhm,Cordova:2016emh}. In particular, in \cite{Cordova:2016emh} we systematically analyzed all unitary short multiplets of superconformal symmetry that contain conserved currents. For 6d SCFTs we found that a non-trivial conserved 2-form current $J_B^{\mu\nu}$ cannot reside in any unitary superconformal multiplet. In short, 6d SCFTs do not allow conserved 2-form currents or continuous 1-form symmetries. 

This observation immediately shows that standard SUSY Yang-Mills gauge theories in six-dimensions (i.e.~theories based on $\CN=(1,0)$ vector multiplets, possibly with hyper multiplet matter) cannot possibly be superconformal. As explained around~\eqref{instcurr}, all such theories possess a conserved 2-form current $J_B^{(2)}$ associated with a $U(1)_B^{(1)}$ instanton number symmetry, which cannot be present in SCFTs. Indeed, it is a well known that gauge theories with conventional Yang-Mills kinetic terms -- and even free Maxwell theory -- are not conformally invariant in six dimensions (see for instance \cite{Jackiw:2011vz,ElShowk:2011gz} and references therein).\footnote{~There are non-standard SUSY gauge theories with 4-derivative kinetic terms, but no Yang-Mills terms \cite{Ivanov:2005kz,Smilga:2006ax}. These theories are not unitary, and are therefore not in conflict with the preceding discussion.} 

From this point of view, it is therefore perhaps surprising that (following~\cite{Seiberg:1996qx}) many interacting 6d SCFTs have been studied via the weakly-coupled description available on their tensor branches, which often contains Yang-Mills gauge fields residing in vector multiplets. In section \ref{ssec:gsintro} below we explain how these theories non-trivially avoid continuous 1-form and 2-group global symmetries, even though they contain Yang-Mills gauge fields.

\subsection{Application 1: Green-Schwarz Terms and 't Hooft Anomalies in 6d SCFTs}\label{ssec:gsintro}

All known interacting 6d SCFTs have a tensor branch -- a moduli space of vacua parametrized by expectation values $\langle \phi \rangle \in \R_+$ of scalars $\phi$ residing in $\CN=(1,0)$ tensor multiplets, which also contain a chiral 2-form gauge field $b^{(2)}$, with self-dual 3-form field strength $h^{(3)} = d b^{(2)} = * h^{(3)}$, and a chiral fermion $\psi^i_\alpha$ in the spin-$\half$ representation of the $SU(2)_R$ symmetry. The largest number of tensor-multiplet scalars that can be activated is called the rank of the tensor branch. For simplicity, we limit our discussion to branches parametrized by a single tensor-multiplet scalar $\phi$, which could either arise as the full tensor branch of rank-1 SCFTs, or as one-dimensional sub-loci of higher-rank tensor branches. The dimension-2 vev $\langle \phi\rangle$ spontaneously breaks conformal invariance, and $\phi$ is the dilaton -- the NG boson of this symmetry breaking. As is typical of NG bosons, the interactions of $\phi$ are highly constrained. This played an essential role in the proof of the 6d SUSY $a$-theorem in \cite{Cordova:2015fha} (see section \ref{ssec:aposintro}). 

In particular, terms in the Lagrangian which would otherwise break conformally symmetry can be rendered conformally invariant by dressing them with suitable powers of the dilaton $\phi$. An important case arises if the low-energy theory on the tensor branch contains dynamical Yang-Mills gauge fields $f^{(2)}_g$, which is the case in many examples. The Yang-Mills kinetic term $\sim \Tr (f^{(2)}_g \wedge * f^{(2)}_g)$ violates conformal symmetry, but dressing it with the dimension-2 dilaton $\phi$ leads to a conformally-invariant kinetic term $\sim \phi \Tr (f^{(2)}_g \wedge * f^{(2)}_g)$. The vev $\langle \phi\rangle \sim {1 \over g_\text{YM}^2}$ then sets the scale of the dimensionful Yang-Mills gauge coupling $g_\text{YM}$. 

Since the coupling of $\phi$ to the Yang-Mills term for $f^{(2)}_g$ restores conformal invariance, it is natural to assume that the couplings of their superpartners restore superconformal invariance. In particular, these couplings must address the puzzle mentioned at the end of section \ref{ssec:no1fmscftint} above: the Yang-Mills theory naturally contains the conserved 2-form instanton current $J_B^{(2)} = i * c_2(f_g^{(2)})$, which is not compatible with superconformal symmetry. This apparent conundrum is resolved by the fact that the dynamical chiral 2-form gauge field $b^{(2)}$ couples to the instanton 2-form current $J_B^{(2)}$ via a GS term
\begin{equation}\label{eq:introgsterm}
S_\text{GS} \sim \int b^{(2)} \wedge * J_B^{(2)} = i \int b^{(2)} \wedge c_2(f_g^{(2)})~,
\end{equation}
which is related by supersymmetry to the dilaton-dressed Yang-Mills term,
\begin{equation}\label{eq:introymterm}
S_\text{YM} \sim \phi \Tr (f^{(2)}_g \wedge * f^{(2)}_g)~. 
\end{equation}
The coupling of $b^{(2)}$ to $J_B^{(2)} $ implies that the latter is no longer a global symmetry current; rather, it has been gauged. The equation of motion of the gauge field $b^{(2)}$ sets $* J_B^{(2)} \sim dh^{(3)}$, which renders $J_B^{(2)}$ trivial as a global 2-form symmetry current (see footnote~\ref{fn:trivcurr}). In the language of conformal operator representations, the fact that $J_B^{(2)}$ is a total derivative makes it a conformal descendant, rather than a primary. Neither an SCFT operator that flows to $h^{(3)}$ on the tensor branch, nor its descendent flowing to $J_B^{(2)}$, need reside in short representations of conformal or superconformal symmetry.\footnote{~It is also worth contemplating the more extreme possibility that there is no well-defined local operator in the SCFT at the origin that flows to the abelian 3-form field strength $h^{(3)}$ on the tensor branch. Loosely speaking, this could be the case if $h^{(3)}$ derives from some intrinsically non-abelian higher-form gauge structure that may be present in the SCFT.} 

The preceding discussion is closely related to anomaly cancellation on tensor branch, which is enforced by a 6d analogue of the GS mechanism~\cite{Green:1984bx, Sagnotti:1992qw,Seiberg:1996qx,Ohmori:2014kda,Intriligator:2014eaa} and involves modifying the Bianchi identity for the field strength $h^{(3)}$ of $b^{(2)}$ as follows,
\begin{equation}\label{eq:gsbianchi}
dh^{(3)} = 0 \quad \Longrightarrow \quad dh^{(3)} = \CI^{(4)}~.
\end{equation}
Here the closed 4-form $\CI^{(4)}$ can contain both dynamical and background fields. Intuitively, one can associate the GS-modified Bianchi identity \eqref{eq:gsbianchi} with a GS term $\sim i \int b^{(2)} \wedge \CI^{(4)}$ in the action, but this is imprecise because $b^{(2)}$ is chiral.\footnote{~As we saw above, precisely such a GS coupling with $\CI^{(4)} \sim i * J_B^{(2)} \sim c_2(f_g^{(2)})$ is responsible for gauging the instanton 1-form symmetry and turning $J_B^{(2)}$ into a conformal descendant.} The fact that $b^{(2)}$ is chiral also implies that \eqref{eq:gsbianchi} gives rise to a GS contribution $\CI^{(8)}_\text{GS}$ to the anomaly polynomial,
\begin{equation}\label{eq:i8gs}
\CI^{(8) }_\text{GS} = {\Omega \over 2} \, \CI^{(4)} \wedge \CI^{(4)}~, \qquad \Omega > 0~.
\end{equation}
The positive constant $\Omega$ defines the Dirac quantization condition for the chiral gauge field~$b^{(2)}$, and we will refer to it as the Dirac pairing.\footnote{~The positivity of the Dirac pairing $\Omega$ is related to the fact that $b^{(2)}$ is a chiral 2-form gauge field. The $\CN=(1,0)$ supergravity multiplet contains an anti-chiral 2-form gauge field whose Dirac pairing is negative.} It is useful to separate the full anomaly polynomial $\CI^{(8)}$ into the GS contribution $\CI^{(8) }_\text{GS}$ in \eqref{eq:i8gs} and a remainder. If the only massless degrees of freedom on the tensor branch are weakly-coupled tensor, vector, and hyper multiplets then this remainder is computed from one-loop box diagrams with external gauge bosons or global symmetry currents, and fermions running in the loop. For this reason we will denote the remainder by $\CI^{(8)}_\text{1-loop}$ and write the full anomaly polynomial as
\begin{equation}\label{anomalyall}
\CI^{(8)} = \CI^{(8)}_\text{1-loop} + \CI^{(8)}_\text{GS}~.
\end{equation}
More generally, the charged massless degrees of freedom that can contribute to $\CI^{(8)}$ may also include strongly-coupled sectors.\footnote{~For instance, this is the case if we exploring the one-dimensional tensor sub-branch of a higher-rank theory.} We will treat such sectors as additional charged matter and include their contribution to the anomaly polynomial in $\CI^{(8)}_\text{1-loop}$. 

We now explain the two anomaly cancellation conditions involving dynamical gauge fields that must be imposed on the tensor branch of any consistent, unitary SCFT: 
\begin{itemize}
\item[\it (i)] {\it Gauge anomaly cancelation:} as is the case for all consistent gauge theories, the term $\CI^{(8)}_\text{gauge} \subset \CI^{(8)}$ in the anomaly polynomial that only depends on the dynamical gauge field $f^{(2)}_g$ must vanish,
\begin{equation}\label{eq:6dgaugezero}
\CI^{(8)}_\text{gauge} = \CI^{(8)}_\text{1-loop, gauge} + \CI^{(8)}_\text{GS, gauge}= 0 \quad \text{in all 6d gauge theories}~.
\end{equation}
The 1-loop gauge anomaly, which arises from charged matter fermions and gauginos (tensor multiplet fermions do not carry gauge charge), takes the general form
\begin{equation}\label{eq:6d1lgauge}
 \CI^{(8)}_\text{1-loop, gauge}  = k_{g^4} \, c_4(f_g^{(2)}) + \half \, k_{(g^2)^2} \, c_2(f_g^{(2)})^2~.
\end{equation}
Since the GS anomaly contribution \eqref{eq:i8gs} is necessarily reducible, the irreducible 1-loop gauge anomaly coefficient must vanish, $k_{g^4} = 0$. However, as emphasized in \cite{Seiberg:1996qx}, the mixed anomaly coefficient is necessarily strictly negative in SCFTs,
\begin{equation}\label{eq:mixedkneg}
k_{(g^2)^2} < 0~,
\end{equation}
and must be canceled using the GS mechanism. This fixes the part of $\CI^{(4)}$ in \eqref{eq:i8gs} that depends on the dynamical gauge field $f_g^{(2)}$,
\begin{equation}\label{eq:i4dynfix}
\CI^{(4)} = \left(- k_{(g^2)^2} \over \Omega\right)^\half c_2(f_g^{(2)}) + \left(\text{background gauge fields}\right)~.
\end{equation}
This expression is meaningful thanks to \eqref{eq:mixedkneg}, which also ensures that the GS and Yang-Mills terms, both of which are proportional to $\sqrt{- k_{(g^2)^2}}$\,, are well defined. 

\item[\it (ii)] {\it Mixed anomaly cancellation:}  As explained in section \ref{ssec:no1fmscftint} above, SCFTs do not allow continuous 1-form symmetries, and a fortiori no continuous 2-group symmetries. However, we also saw in section \ref{ssec:mixanom2gpintro} that such 2-group symmetries necessarily arise whenever the anomaly polynomial of a 6d gauge theory contains mixed gauge-global terms $\CI^{(8)}_\text{mixed} = c_2(f_g^{(2)}) \wedge \left(\text{background gauge fields}\right)$, where the background gauge fields are associated with 0-form global symmetries (e.g.~a flavor symmetry as in \eqref{eq:Imixedex}, or Poincar\'e symmetry as in \eqref{Imixedeg}). Together these two facts lead to the inexorable conclusion that all such mixed gauge-global anomalies must vanish in 6d SCFTs,\footnote{~Here we assume that the 0-form symmetries participating in $\CI_\text{mixed}^{(8)}$ are exact symmetries of the full SCFT, rather than accidental symmetries that only emerge at low energies on the tensor branch. This is true for Poincar\'e symmetry, but may not hold for all flavor symmetries. A candidate example of a 6d SCFT with an emergent flavor symmetry on its tensor branch was discussed in \cite{Ohmori:2015pia}.}
\begin{equation}\label{nomixedanoms}
\CI^{(8)}_\text{mixed} = \CI^{(8)}_\text{1-loop, mixed} + \CI^{(8)}_\text{GS, mixed}= 0 \quad \text{in all 6d SCFTs}~.
\end{equation}
The 1-loop mixed anomaly only arises from charged matter fermions and gauginos and takes the general form
\begin{equation}\label{eq:i1lmixed}
\CI^{(8)}_\text{1-loop, mixed} = k_{g^2 G^2} \, c_2(f_g^{(2)}) \, c_2(F_G^{(2)}) + {k_{g^2 \SP^2} \over 24} \, c_2(f_g^{(2)}) \, p_1(T) \equiv c_2(f_g^{(2)}) \wedge X^{(4)}_\text{global}~.
\end{equation}
If the Lie algebra of $G^{(0)}$ is not compact and simple, but rather a direct sum of compact, simple (or abelian) Lie algebras, then the term $\sim c_2(F_G^{(2)})$ in \eqref{eq:i1lmixed} is replaced by a suitable sum of such terms. For instance, this happens if the SCFT has a flavor symmetry that commutes with SUSY,  in addition to the unavoidable $SU(2)_R$ symmetry.

In order to satisfy \eqref{nomixedanoms}, the reducible mixed gauge-global anomalies in \eqref{eq:i1lmixed} must be cancelled by the GS contribution $\CI^{(8)}_\text{GS, mixed}$. Together with \eqref{eq:i4dynfix}, this completely determines $\CI^{(4)}$, and hence the associated GS term $\sim i \int b^{(2)} \wedge \CI^{(4)}$,
\begin{equation}
\CI^{(4)} = \left(- k_{(g^2)^2} \over \Omega\right)^\half \left( c_2(f_g^{(2)}) + {1 \over k_{(g^2)^2} } X^{(4)}_\text{global} \right) ~.
\end{equation}
This in turn fixes the full GS contribution $\CI^{(8)}_\text{GS} = {\Omega \over 2} \, \CI^{(4) }\wedge \CI^{(4)}$ in \eqref{eq:i8gs} to the anomaly polynomial. Taking into account the vanishing of $\CI^{(8)}_\text{gauge}$ and $\CI^{(8)}_\text{mixed}$ from \eqref{eq:6dgaugezero} and \eqref{nomixedanoms}, we find that the full anomaly polynomial in \eqref{anomalyall} reduces to 
\begin{equation}\label{eq:fullgloabli8}
\begin{split}
\CI^{(8)} = \CI^{(8)}_\text{global} = ~ & \CI^{(8)}_\text{1-loop, global} - {1 \over 2 k_{(g^2)^2} } X^{(4)}_\text{global} \wedge X^{(4)}_\text{global} \\
=~& \CI^{(8)}_\text{1-loop, global} - {1 \over 2 k_{(g^2)^2} } \left(k_{g^2 G^2}  \, c_2(F_G^{(2)}) + {k_{g^2 \SP^2} \over 24} \, p_1(T) \right)^2~.
\end{split}
\end{equation}
Since there is no 2-group symmetry, all terms in $\CI^{(8)}_\text{global}$ represent meaningful 't Hooft anomalies of the SCFT (see section \ref{ssec:2gphooft}). All terms on the right-hand side of \eqref{eq:fullgloabli8} can be computed from anomaly coefficients in $\CI^{(8)}_\text{1-loop}$, i.e.~from the gauge and global symmetry quantum numbers of massless degrees of freedom on the tensor branch. Note in particular that the Dirac pairing $\Omega$ cancels out of the final expression for $\CI^{(8)}$, even though it appeared throughout the preceding discussion. 
\end{itemize}

The fact that setting all mixed gauge-global anomalies $\CI^{(8)}_\text{mixed} = 0$ to zero fixes the~'t~Hooft anomalies of 6d SCFTs according to \eqref{eq:fullgloabli8} was understood in~\cite{Ohmori:2014kda}, where this procedure was applied to many examples and subjected to numerous consistency checks. However, the physical reason why $\CI^{(8)}_\text{mixed} = 0$ should be imposed in the first place has remained mysterious. Here we see that this condition need only be imposed if the gauge theory under consideration descends from an SCFT in the UV, since the latter cannot have continuous 1-form or 2-group global symmetries. By contrast, the non-conformal gauge theories that arise in the IR of many 6d LSTs can have $\CI^{(8)}_\text{mixed} \neq 0$, and thus do enjoy 2-group symmetry (at the very least as an emergent symmetry in the deep IR, and possibly beyond, see sections \ref{ssec:introlst} and~\ref{sec:LSTs}). This shows that no completely general consistency condition, nor supersymmetry or UV completeness alone, can force $\CI^{(8)}_\text{mixed} = 0$. Rather, this condition follows from the stronger assumption of a superconformal UV fixed point.

\subsection{Application 2: Positivity of the $a$-Type Weyl Anomaly in 6d SCFTs}\label{ssec:aposintro}

We now apply our improved understanding of 't Hooft anomalies in 6d SCFT in the context of the $a$-theorem for this class of theories.  The Weyl anomalies of any 6d CFT are captured by the (scheme-independent terms in) the anomalous trace of the stress tensor in the presence of a background metric. In six dimensions \cite{Deser:1993yx}, this trace takes the form
\begin{equation}
\langle T_\mu ^\mu \rangle \sim a E_6+\sum _{i=1}^3 c_i I_i \ + \ (\text{scheme dependent})~,\label{Tanom}
\end{equation}
where~$E_6$ is the six-dimensional Euler density and $I_{1,2,3}$ are weight-6 Weyl invariants. All four Weyl anomaly coefficients $a, c_i$ are interesting observables, but the $a$-anomaly plays a privileged role -- primarily due to its conjectured monotonicity properties under RG flows. Following \cite{Cardy:1988cwa}, this conjecture -- commonly referred to as the $a$-theorem, despite the absence of a general proof -- states that the $a$-anomaly decreases for unitary RG flows connecting conformal fixed points in the UV and IR,\footnote{~The $a$-theorem is an established result in 2d \cite{Zamolodchikov:1986gt} (where it is known as the $c$-theorem) and in 4d \cite{Komargodski:2011vj}.}
\begin{equation}\label{eq:athm}
\Delta a = a_\text{UV} - a_\text{IR} > 0~.
\end{equation}
If we use the $a$-anomaly as a proxy for the number of effective degrees of freedom in a CFT, the inequality \eqref{eq:athm} captures the conventional RG intuition that this number should decrease as we integrate out degrees of freedom along an RG flow. While a proof of the 6d $a$-theorem for general, non-supersymmetric RG flows remains elusive,\footnote{~See \cite{Elvang:2012st} for an attempt at a proof along the lines of \cite{Komargodski:2011vj}, and \cite{Kundu:2019zsl} for an attempt via the conformal bootstrap.} it has been proved for all RG flows onto the tensor branch of 6d SCFTs \cite{Cordova:2015vwa,Cordova:2015fha}, and verified for some examples of flows onto Higgs branches \cite{Heckman:2015axa,Apruzzi:2017nck}. As was already mentioned above, such moduli space RG flows are the only SUSY-preserving flows out of 6d SCFTs~\cite{Louis:2015mka,Cordova:2016xhm}.

Another property that the $a$-anomaly is expected to have based on its interpretation as counting effective degrees of freedom is that it should be positive in unitary CFTs,
\begin{equation}\label{eq:apos}
a > 0~.
\end{equation}
This inequality is logically distinct from the $a$-theorem inequality \eqref{eq:athm}, but  it is closely related. For instance, if there is an RG flow from a given CFT to a gapped theory, then $a_\text{IR} = 0$ and the $a$-theorem~\eqref{eq:athm} implies that the $a$-anomaly of the CFT is positive, as in~\eqref{eq:apos}. However, not all CFTs necessarily admit such flows to gapped phases. 

In \cite{Cordova:2015fha} we studied the $a$-theorem for RG flows of a 6d $\CN=(1,0)$ SCFT onto its tensor branch. In addition to proving \eqref{eq:athm} in this context, we argued for a general anomaly multiplet relation expressing the $a$-anomaly of the SCFT in terms of its 't Hooft anomalies for $SU(2)_R$ and Poincar\'e symmetries,\footnote{~Similar $\CN=(1,0)$ anomaly multiplet relations for the $c_i$-anomalies were proposed in \cite{Yankielowicz:2017xkf,Beccaria:2017dmw} and proved  in \cite{Cordova:2019wns}. The corresponding results for $\CN=(2,0)$ theories were obtained in \cite{Cordova:2015vwa}.}
\begin{equation}\label{ais}
a_\text{SCFT}= \frac{16}{7} \, \left(\alpha -  \beta +  \gamma \right)+ \frac{6}{7} \, \delta~.
\end{equation}
For the purpose of this discussion we follow the notation and conventions of \cite{Cordova:2015fha}: the anomaly coefficients $\alpha, \beta, \gamma, \delta$ are defined as follows,
\begin{equation}\label{anomalyI}
\CI^{(8)}_\text{global} \quad \supset  \quad {1 \over 4!} \Big(\alpha c_2^2(R) + \beta c_2(R) p_1(T) + \gamma p_1^2(T) + \delta p_2(T)\Big)~,
\end{equation}
where $c_2(R)  \equiv c_2\big(F_{SU(2)_R}^{(2)}\big)$ is the second Chern class for the $SU(2)_R$ background gauge field, while $p_{1}(T)$ and $p_{2}(T)$ are the first and second Pontryagin classes of the tangent bundle. The overall normalization of the $a$-anomaly is chosen such that a free $\CN=(2,0)$ tensor multiplet (consisting of one $\CN=(1,0)$ tensor and one $\CN=(1,0)$ hypermultiplet) has $a = 1$. Table~\ref{known} lists the  't Hooft anomaly coefficients $\alpha, \beta, \gamma, \delta$, as well as the value for the $a$-anomaly computed using \eqref{ais}, for several $ \CN=(1,0)$ and $\CN=(2,0)$ theories.  

\bigskip

\begin{table}[h]
\centering
\begin{tabular}{!{\VRule[1pt]}c!{\VRule[1pt]}c!{\VRule[1pt]}c!{\VRule[1pt]}c!{\VRule[1pt]} c!{\VRule[1pt]}c!{\VRule[1pt]}}
\specialrule{1.2pt}{0pt}{0pt}
{\bf Theory} & $\bf \alpha$ &  $\bf \beta$ & $\bf \gamma$ & $\bf \delta$ & $a$ \rule{0pt}{2.6ex}\rule[-1.4ex]{0pt}{0pt} \\
\specialrule{1.2pt}{0pt}{0pt}
\multirow{2}{*}{ $\CN=(1,0)$ hyper multiplet}& \multirow{2}{*}{$0$} &  \multirow{2}{*}{$0$} & \multirow{2}{*}{$7 \over 240$} & \multirow{2}{*}{$-{1 \over 60}$} & \multirow{2}{*}{$11 \over 210$}\\
 &  & & & & \\
\hline
\multirow{2}{*}{$\CN=(1,0)$ tensor multiplet}& \multirow{2}{*}{$1$} &  \multirow{2}{*}{$\half$} & \multirow{2}{*}{$23 \over 240$} & \multirow{2}{*}{$-{29 \over 60}$} & \multirow{2}{*}{$199 \over 210$}\\
 &  & & & & \\
 \hline
 \multirow{2}{*}{$\CN=(1,0)$ vector multiplet}& \multirow{2}{*}{$-1$} &  \multirow{2}{*}{$- \half $} & \multirow{2}{*}{$-{7 \over 240}$} & \multirow{2}{*}{$1 \over 60$}& \multirow{2}{*}{``\,$- {251 \over 210}$\, ''}\\
 &  & & & & \\
\hline
 \multirow{2}{*}{$\CN=(2,0)$ theory of ADE type $\frak{g}$}& \multirow{2}{*}{$h^\vee _{\frak{g}}d_{\frak{g}}+r_{\frak{g}}$} &  \multirow{2}{*}{$\half r_{\frak{g}} $} & \multirow{2}{*}{$\frac{1}{8}r_{\frak{g}}$} & \multirow{2}{*}{$-\half r_{\frak{g}}$}& \multirow{2}{*}{$\frac{16}{7}h_{\frak{g}}^\vee d_{\frak{g}}+r_{\frak{g}}$}\\
 &  & & & & \\
\specialrule{1.2pt}{0pt}{0pt}
 \end{tabular}
\caption{'t Hooft and $a$-anomalies for some SUSY theories. The negative entry for the $a$-anomaly of the $\CN=(1,0)$ vector multiplet should not be taken at face value (hence the double quotes) and is further discussed in the text. The $\CN=(2,0)$ theories are associated with an ADE Lie algebra $\frak g$, with rank $r_\frak{g}$, dimension $d_\frak{g}$, and dual Coxeter number $h^\vee_\frak{g}$. }
\label{known}
\end{table}
\noindent

\bigskip

A surprising entry in Table~\ref{known} is the $a$-anomaly for a free $\CN=(1,0)$ vector multiplet, computed by substituting its 't Hooft anomaly coefficients $\alpha, \beta, \gamma, \delta$ into the superconformal anomaly multiplet relation \eqref{ais}. The resulting number is negative, $a_\text{vector} = - {251 \over 210}$. As emphasized in \cite{Cordova:2015fha} this result is puzzling but not a contradiction: since the free $\CN=(1,0)$ vector multiplet is not an SCFT, there is no immediate meaning to the $a$-anomaly (or any other Weyl anomaly), nor to the anomaly multiplet relation \eqref{ais}.\footnote{~A direct interpretation of the formula $a_\text{vector} = - {251 \over 210}$ was pointed out in \cite{Beccaria:2015uta}: it correctly computes the actual $a$-type Weyl anomaly of a superconformal, but non-unitary, theory constructed using a free $\CN=(1,0)$ vector multiplet with 4-derivative kinetic terms, but no conventional 2-derivative kinetic terms. Since the~'t~Hooft anomalies $\alpha, \beta, \gamma, \delta$ of this theory only depend on the massless spectrum (not the kinetic terms), they agree with those computed for an ordinary free vector multiplet with Maxwell kinetic term. This observation will not play a role in our discussion, since we only consider unitary theories.} More concerning is the fact that many 6d SCFTs possess tensor branches with Yang-Mills gauge fields residing in $\CN=(1,0)$ vector multiplets in the IR. If the massless spectrum on the tensor branch consists of $n_H$ hyper, $n_T$ tensor, and $n_V$ vector multiplets, Table~\ref{known} implies that the naive low-energy contributions of these fields to the $a$-anomaly of the UV SCFT takes the form
\begin{equation}\label{eq:ascft}
a_\text{SCFT} = a_\text{IR} + \Delta a~, \qquad a_\text{IR} = \frac{11}{210}n_H+\frac{199}{210}n_T-\frac{251}{210}n_V~.
\end{equation} 
If the IR theory contains sufficiently many vector multiplets, then $a_\text{IR}$ may be negative. In this case the $a$-theorem $\Delta a > 0$ is not sufficient to guarantee that $a_\text{SCFT}>0$. 

In fact, this phenomenon is rather generic. Consider the $\CN=(1,0)$ SCFTs that can be engineered in type IIA string theory using two NS 5-branes suspended inside a stack of $N \geq 2$ D6-branes \cite{Brunner:1997gk, Blum:1997mm,Brunner:1997gf,Hanany:1997gh}. Separating the NS 5-branes along the D6-branes opens up a rank-1 tensor branch.\footnote{~We ignore the decoupled center of mass mode of the NS 5-branes, which does not affect our argument.} The massless spectrum on the tensor branch consists of one tensor multiplet ($n_T = 1$), an $SU(N)$ Yang-Mills theory with $n_V = N^2 - 1$, and $N_f = 2N$ fundamental hyper multiplets, so that $n_H = 2N^2$.\footnote{~Precisely this number of hypers is generically needed to cancel the irreducible quartic gauge anomaly, i.e.~to ensure that $k_{g^4} = 0$.} Substituting into \eqref{eq:ascft} we find that the contribution of these massless fields to the $a$-anomaly of the SCFT is
\begin{equation}
a_\text{IR} = {1 \over 210} \left(199 + 22 N^2 - 251 (N^2 - 1) \right) = {1 \over 210} \left(450 -229 N^2 \right) < 0~.
\end{equation}
An intuitive reason why this phenomenon occurs quite frequently is that gauge-anomaly cancellation limits the amount of hyper-multiplet matter for a given gauge group, so that the negative vector-multiplet contribution to $a_\text{IR}$ in \eqref{eq:ascft} is often significant.  

As the preceding example shows, it is not sufficient to rely on the $a$-theorem inequality $\Delta a > 0$ to prove that $a_\text{SCFT} = a_\text{IR} + \Delta a > 0$, since $a_\text{IR}$ may be negative. Instead, we must control the actual magnitude of $\Delta a$ relative to $a_\text{IR}$. Note that $a_\text{IR}$ can only be negative if there are vector multiplets on the tensor branch.\footnote{~In SCFTs whose low-energy description on the tensor branch only contains tensor and hyper multiplets, such as the $\CN=(2,0)$ theories or the small $E_8$ instanton SCFTs, $a_\text{IR} > 0$ so that $a_\text{SCFT} > 0$ does indeed follow from the $a$-theorem \cite{Cordova:2015fha}.} As we saw in section \ref{ssec:gsintro} above, this is precisely the situation in which the absence of 2-group global symmetries (enforced via the absence of mixed gauge-global anomalies) allows us to compute the full 't Hooft anomalies $\CI^{(8)}_\text{global}$ using only the low-energy spectrum on the tensor branch. If we apply the general formula \eqref{eq:fullgloabli8} with $G^{(0)} \rightarrow SU(2)_R^{(0)}$ and $c_2\big(F_G^{(2)}\big) \rightarrow c_2(R)$, $k_{g^2 G^2} \rightarrow k_{g^2 R^2}$, we find that
\begin{equation}\label{eq:rpgs}
\CI^{(8)}_\text{global} = \CI^{(8)}_\text{1-loop, global} - {1 \over 2 k_{(g^2)^2} } \left(k_{g^2 R^2}  \, c_2(R) + {k_{g^2 \SP^2} \over 24} \, p_1(T) \right)^2~.
\end{equation}
Applying the formula \eqref{ais} for the $a$-anomaly to $\CI^{(8)}_\text{1-loop, global}$, we find the formula for $a_\text{IR}$ in \eqref{eq:ascft}. Consequently, substituting the GS contribution on the right-hand side of \eqref{eq:rpgs} into $\eqref{ais}$ computes $\Delta a$,
\begin{equation}\label{eq:deltagsintro}
\Delta a =  -\frac{192}{7 k_{(g^2)^2}} \left(k_{g^2 R^2} - {k_{g^2 \SP^2} \over 24}  \right)^2~.
\end{equation}
Note that this expression is positive thanks to \eqref{eq:mixedkneg}. It is not an accident that both the GS terms in \eqref{eq:rpgs} and $\Delta a$ in \eqref{eq:deltagsintro} are very similar perfect squares (note, however, the relative minus sign), since both are quadratically related to GS terms in the Lagrangian~\cite{Cordova:2015fha}.

The explicit formula \eqref{eq:deltagsintro} allows us to compute $\Delta a$ from the massless spectrum on the tensor branch precisely when there are IR gauge fields, where we must prove  that $\Delta a$ is large enough to overwhelm a potentially negative value of $a_\text{IR}$ caused by vector-multiplet contributions. In section \ref{sec:scfts} we prove this using \eqref{eq:deltagsintro} as well as other constraints, such as conventional gauge-anomaly cancelation. As in our proof of the $a$-theorem~\cite{Cordova:2015fha}, our argument that $a_\text{SCFT} > 0$ only relies on general consistency conditions and does not depend on any explicit enumeration or classification of 6d SCFTs.

\section{Further Aspects of 6d Theories with 2-Group Global Symmetries}\label{sec:otheraspects}

\subsection{2-Group Background Gauge Fields from Mixed Gauge-Global Anomalies}\label{ssec:moreImixed}

Consider a 6d Yang-Mills gauge theory with gauge group $g^{(0)}$ and field strength $f_g^{(2)}$. This theory has a $U(1)_B^{(1)}$ instanton 1-form symmetry associated with the conserved 2-form current in \eqref{instcurr}, which we repeat here,
\begin{equation}\label{instcurr2}
J^{(2)}_B = i*c_2(f^{(2)})~, \qquad c_2(f^{(2)}) =  \frac{1}{8 \pi^2}\Tr \Big(f^{(2)} \wedge f^{(2)}\Big)~.
\end{equation}
As in \eqref{BwedgeJ}, the coupling of $J_B^{(2)}$ to the corresponding background 2-form gauge field $B^{(2)}$ takes the form
\begin{equation}\label{Bcoupling}
S \supset \int B^{(2)} \wedge * J_B^{(2)} = i \int B^{(2)}\wedge c_2(f^{(2)}_g)~.
\end{equation}
The conservation equation $d*J_B^{(2)} = 0$ corresponds to invariance under background 1-form gauge transformations \eqref{eq:bgt}, $B^{(2)} \rightarrow B^{(2)}+d\Lambda _B^{(1)}$.

We next turn to the effect of anomalies for gauge and global 0-form symmetries, encoded by an anomaly polynomial $\CI^{(8)}$ via the descent procedure. If we use $\delta$ to collectively denote all dynamical and background gauge variations, then the total gauge variation $\delta S_\text{eff}$ of the effective action\footnote{~We permit ourselves an abuse of language and define the effective action $S_\text{eff}$ to be obtained by integrating out all dynamical fields in the path integral, except for dynamical gauge fields. Then $S_\text{eff}$ is a (typically highly non-local) functional of both dynamical and background gauge fields, and we can discuss its (non-) invariance under the associated gauge transformations.}  is given by the following formulas,
\begin{equation}\label{descent}
\delta S_\text{eff} = 2\pi i \int \CI ^{(6)}~, \qquad \CI ^{(8)}=d\CI ^{(7)}~, \qquad  \delta \CI ^{(7)}=d\CI ^{(6)}~.
\end{equation} 
Let us apply these formulas to the case of mixed-global anomalies involving the dynamical gauge group $g^{(0)}$, and various 0-form symmetries. As in the introduction, we consider 0-form flavor symmetries $G^{(0)}$ and Poincar\'e symmetry $\SP^{(0)}$ (see~\eqref{eq:Imixedex} and~\eqref{Imixedeg}),
\begin{equation}\label{mixedcoeffs}
\CI^{(8)}_\text{mixed} = c_2(f^{(2)}_g) \wedge \left(k_{g^2G^2} c_2(F^{(2)}_G)+\frac{k_{g^2\SP^2}}{24}p_1(T)\right)~.
\end{equation}
We are free to add 6d local counterterms to the action (and hence to $S_\text{eff}$), whose gauge variation may contribute to \eqref{descent}. (The freedom of adding local 6d counterterms is an inherent ambiguity in the descent procedure.) Two important counterterms related to the mixed anomalies in \eqref{mixedcoeffs} are proportional to $\int \text{CS}(a_g^{(1)}) \wedge  \text{CS}(A_g^{(1)})$ and $\int \text{CS}(a_g^{(1)}) \wedge  \text{CS}(\omega^{(1)})$, where the Chern-Simons terms were defined in \eqref{eq:csdef}, and their change under gauge transformations can be found using \eqref{eq:csvar}. By tuning the coefficients of these counterterms, we can make the mixed anomalies \eqref{mixedcoeffs} appear either in dynamical $g^{(0)}$ gauge transformations, or background $G^{(0)}$ gauge transformations and local Lorentz transformations. 

Consistency of the path integral of $e^{-S_\text{eff}}$ over the dynamical gauge field $a_g^{(1)}$, modulo $g^{(0)}$ gauge transformations, requires that these gauge transformations be completely free of anomalies -- even in the presence of background fields. This uniquely fixes the coefficients of the counterterms mentioned above (see~\cite{Cordova:2018cvg} for a detailed discussion of the 4d case), and it leads to an unambiguous variation of $S_\text{eff}$ under background gauge transformations of $A_G^{(1)}$, and under local Lorentz transformations (parametrized by $\lambda_G^{(0)}$ and $\theta^{(0)}$, see \eqref{eq:aegt}),
\begin{equation}\label{ggGGvariation}
\delta S_\text{eff} = {i \over 4 \pi} \int \left( k_{g^2 G^2} \Tr \left(\lambda_G^{(0)}  d A_G^{(1)}\right) - {k_{g^2 \SP^2} \over 24} \tr \left(\theta^{(0)} d \omega^{(1)}\right)\right) \wedge c_2(f_g^{(2)})~.
\end{equation}
Note that the effective action shifts by the operator $c_2(f_g^{(2)})$ constructed out of the dynamical gauge fields under background $G^{(0)}$ gauge transformations and local Lorentz transformations. If these transformations are associated with honest symmetries, $S_\text{eff}$ should be invariant under these transformations (modulo possible c-number 't Hooft anomalies). Fortuitously, $c_2(f_g^{(2)})$ is sourced by the 2-form background gauge field $B^{(2)}$ via \eqref{Bcoupling}. We can therefore cancel the variation in \eqref{ggGGvariation} by postulating the following transformation rule for $B^{(2)}$ under all background gauge transformations, which is exactly of the GS form \eqref{eq:Bshiftsintro},
\begin{equation}\label{eq:Bshifts2}
B^{(2)} \rightarrow B^{(2)} +d\Lambda _B^{(1)}  +\frac{\hat \kappa _G}{4\pi}\Tr\left( \lambda _G^{(0)}dA_G^{(1)}\right)+\frac{\hat \kappa _\SP}{16\pi} \tr \left(\theta ^{(0)}d\omega ^{(1)}\right)~,
\end{equation}
with
\begin{equation}\label{eq:2gpconstanom}
\hat \kappa _G = - k_{g^2 G^2}~, \qquad \hat \kappa _\SP = {k_{g^2 \SP^2} \over 6}~.
\end{equation}
As was explained in section \ref{ssec:2gpreview}, this transformation law is one of several equivalent ways of stating the fact that the 0-form symmetries $G^{(0)}$ and $\SP^{(0)}$ join with the instanton 1-form symmetry $U(1)_B^{(1)}$ to form a non-trivial 2-group (i.e.~a twisted product symmetry, as in \eqref{eq:twistedsym}). The 2-group structure constants $\hat \kappa_G$ and $\hat \kappa_\SP$ are determined by the mixed gauge-global anomaly coefficients according to \eqref{eq:2gpconstanom}. All other consequences of 2-group symmetry can be derived from \eqref{eq:2gpconstanom}. For instance, together with footnote \ref{currnorm} and the invariance of $S_\text{eff}$ under background gauge transformations, we can use it to derive the 2-group non-conservation equations \eqref{modcons}. 

\subsection{Anomaly Polynomial and 2-Group Structure Constants for Free Fermions}

Throughout the remainder of this paper, we will need the anomaly polynomial for free fermions transforming under various gauge and global symmetries. An immediate application is that mixed gauge-global anomaly coefficients determine the 2-group structure constants, as in \eqref{eq:2gpconstanom}, and we compute them here.

The anomaly polynomial for spin-$\half$ fermions can be extracted from \cite{AlvarezGaume:1983ig,AlvarezGaume:1984dr}. Consider a chiral spin-$\half$ (i.e.~a left-handed Weyl) fermion in 6d, which transforms in a representation $\rho_g$ (of dimension $|\rho_g|$) of the dynamical gauge group $g^{(0)}$, and in a representation $\rho_G$ (of dimension $|\rho_G|$) of the flavor symmetry $G^{(0)}$. Altogether, the fermion transforms in the product representation $\rho_g \otimes \rho_G$, which has dimension $|\rho_g| |\rho_G|$. Its anomaly polynomial  is then given by
\begin{equation}\label{eq:i8chif}
\CI^{(8)}_\text{chiral} (\rho_g, \rho_G) =  \Tr_{\rho_g \otimes \rho_G} \left( \hat A \, \exp \left({i \over 2\pi} \left( f^{(2)}_g \big|_{\rho_g \otimes \1_G} + F^{(2)}_G \big|_{\1_g \otimes \rho_G}\right) \right)\right) \bigg|_\text{8-form}~,
\end{equation}
where we are instructed to expand the expression inside the trace and project it onto 8-forms. The $\hat A$-genus can be expanded in Pontryagin classes $p_i(T)$ of the tangent bundle; the first few terms of the  are given by
\begin{equation}
 \hat A(R) = 1 - {1 \over 24} p_1(T) + {1 \over 2^3 6!} \Big( 7 p_1(T)^2 - 4 p_2(T)\Big) + \cdots~.
\end{equation} 
Substituting into \eqref{eq:i8chif} and carrying out this procedure, we find that 
 \begin{equation}\label{sixdanomg}
\begin{split}
\CI^{(8)}_\text{chiral} (\rho_g, \rho_G) =~& \frac{|\rho _G|}{4! (2\pi)^4}\Tr _{\rho _g} (f_g^{(2)})^4+\frac{|\rho _g|}{4!(2\pi)^4}\Tr _{\rho _G} (F_G^{(2)})^4+\frac{1}{2^2(2\pi)^4}\Tr _{\rho _g}(f_g^{(2)})^2 \, \Tr _{\rho _G}(F_G^{(2)})^2 \\
 &+\frac{1}{48(2\pi )^2} \, p_1(T) \, \left(|\rho _g|\Tr _{\rho _G}(F_G^{(2)})^2+|\rho _G|\Tr _{\rho _g}(f_g^{(2)})^2\right) \\
 & +\frac{|\rho _G||\rho _g|}{2^36!} \Big(7p_1(T)^2-4p_2(T)\Big).
 \end{split}
\end{equation}
The corresponding result for an anti-chiral spin-$\half$ (i.e.~for a right-handed Weyl) fermion in 6d is identical up to an overall sign,
\begin{equation}\label{eq:antichii}
\CI^{(8)}_\text{anti-chiral} (\rho_g, \rho_G) = -\CI^{(8)}_\text{chiral} (\rho_g, \rho_G)~.
\end{equation}

We can rewrite the mixed anomaly terms in \eqref{sixdanomg} in terms of second Chern classes, which are normalized as in~\eqref{instcurr}, 
\begin{equation}\label{eq:casimir}
\half \Tr _{\rho_g} \left(\frac{f_g^{(2)}}{2\pi}\right)^2=T_2(\rho_g) c_2(f_g^{(2)})~, \qquad \half \Tr _{\rho_G} \left(\frac{F_G^{(2)}}{2\pi}\right)^2=T_2(\rho_G) c_2(F_G^{(2)})~.
\end{equation}
Here the non-negative integer $T_2(\rho) \in \Z_{\geq 0}$ is the quadratic (or Dynkin) index of the representation $\rho$,\footnote{~The index $T_2(\rho)$ is related to the quadratic Casimir $C_2(\rho)$ via $T_2(\rho) = {C_2(\rho) \text{dim}(\rho) \over \text{dim}(g)}$, with $C_2(\rho)$ normalized so that $C_2(\text{adj}) = 2 h_g^\vee$. Note that $T_2(\rho)$ is the index of the 4d Dirac operator in the representation $\rho$. Thus it counts the number of zero modes for a chiral fermion in the  representation $\rho$ in the presence of a unit instanton in 4d.} normalized so that $T_2(\text{adj})=2h^\vee$ for the adjoint representation, with $h^\vee$ the dual Coxeter number (see for instance \cite{DiFrancesco:1997nk}). The fundamental representations $\square$ of $SU(N)$, $SO(N \geq 5)$, and $Sp(N)$ (of dimensions $N$, $N$, and $2N$, respectively) have $T_2(\square) = 1$, $T_2(\square) = 2$, and $T_2(\square) = 1$ respectively. Comparing~\eqref{sixdanomg} with \eqref{mixedcoeffs}, we see that the mixed gauge-global anomaly coefficients for a chiral fermion are given by 
 \begin{equation}\label{ffanomalycoeffs}
\quad k_{g^2G^2}=T_2(\rho _g)T_2(\rho _G)~, \qquad k_{g^2\SP^2}=T_2(\rho _g) |\rho _G|~.
 \end{equation} 
Both of these anomaly coefficients are integers, as advertised in \eqref{eq:kgGquant} and \eqref{eq:kgPquant}. Substituting into \eqref{eq:2gpconstanom}, we find that the 2-group structure constants resulting from these mixed gauge-global anomalies are
\begin{equation}\label{eq:ffkgkp}
\hat \kappa_G = - T_2(\rho _g)T_2(\rho _G)~, \qquad \hat \kappa_\SP = {T_2(\rho _g) |\rho _G| \over 6}~.
\end{equation}
Note that $\hat \kappa_G$ is quantized in accordance with \eqref{kappaquant}, while this is not obviously the case for $\hat\kappa_\SP$. See section \ref{ssec:quantcond} below for further discussion of this point.

\subsection{Quantization of the Poincar\'e 2-Group Structure Constant}\label{ssec:quantcond}

Here we would like to explain why the 2-group structure constant  $\hat \kappa_\SP$ is always an integer, even though this is not obvious from \eqref{eq:2gpconstanom}, which we repeat here,
\begin{equation}
\hat \kappa_\SP = {k_{g^2 \SP^2} \over 6}~.
\end{equation}
Even though $k_{g^2 \SP^2}$ is always an integer, this integer is not manifestly divisible by $6$. In order to understand the physics underlying the resolution of this puzzle it is sufficient to consider a simple example: take the gauge group to be abelian $g^{(0)} =  U(1)$, the flavor symmetry $G^{(0)}$ to be absent, and the only charged matter to be chiral fermions $\psi_i$ of $U(1)$ charge $q_i \in \Z$ or anti-chiral fermions $\t \psi_{\t i}$ of $U(1)$ charge $\t q_{\t i} \in \Z$. It is straightforward to extend the discussion to non-abelian gauge groups (e.g.~by considering their Cartan subgroups), to include the flavor symmetry $G^{(0)}$, and to generalize the arguments to general anomaly polynomials, beyond free fermion examples. A nearly identical puzzle about the integrality of Poincar\'e 2-group coefficients arises in 4d; see \cite{Cordova:2018cvg} for a detailed discussion of that puzzle, without simplifying assumptions. 

If we repeat the derivation leading to \eqref{sixdanomg} and \eqref{eq:antichii} for our abelian example, we find that the anomaly polynomial is given by\footnote{~In the abelian example discussed here we depart from our convention of taking the field strength $f_g^{(2)}$ to be anti-hermitian and take it to be real instead. This explains the missing factors of $i$ in the exponents of \eqref{eq:abap} relative to those in \eqref{eq:i8chif}.}
\begin{equation}\label{eq:abap}
\CI^{(8)} = \hat A \bigg(\sum_i e^{q_i f_g^{(2)} \over 2 \pi}-  \sum_{\t i} e^{\t q_{\t i} f_g^{(2)} \over 2 \pi}\bigg)\bigg|_\text{8-form}~.
\end{equation}
We focus on the gauge and mixed parts of $\CI^{(8)}$, 
\begin{equation}
\begin{split}
& \CI^{(8)}_\text{gauge} = {1 \over 4! (2\pi)^4} \bigg(\sum_i q_i^4 - \sum_{\t i} (\t q_{\t i})^4\bigg) (f_g^{(2)})^4~, \\
& \CI^{(8)}_\text{mixed} =  \bigg(\sum_i q_i^2 - \sum_{\t i} (\t q_{\t i})^2\bigg)  {1 \over 8 \pi^2} f_g^{(2)} \wedge f^{(2)}_g \wedge \left( - {p_1(T) \over 24}\right)~.
\end{split}
\end{equation}
Identifying ${1 \over 4\pi^2} f^{(2)}_g \wedge f^{(2)}_g$ as the properly normalized $U(1)$ instanton density (see the discussion below \eqref{eq:ckjdef}), we conclude that the mixed anomaly coefficient $k_{g^2 \SP^2}$ we are after is given by
\begin{equation}\label{qe:u1mixedan}
k_{g^2 \SP^2} = -  \half \bigg(\sum_i q_i^2 - \sum_{\t i} (\t q_{\t i})^2\bigg)~,
\end{equation}
while the pure $U(1)$ gauge anomaly coefficient $k_{g^4}$ must be set to zero, 
\begin{equation}\label{eq:u1gac}
k_{g^4} \sim \sum_i q_i^4 - \sum_{\t i} (\t q_{\t i})^4 = 0~.
\end{equation}

We will now show that the divisibility of the mixed anomaly coefficient $k_{g^2 \SP^2}$ in \eqref{qe:u1mixedan} by 6 follows from the $U(1)$ gauge-anomaly cancellation condition \eqref{eq:u1gac}. To see this, it is sufficient to observe that any integer $n^4 \equiv n^2$ modulo 2, 3, 4, and hence also modulo 12. In equations, $n^4 \equiv n^2 (\text{mod } 12)$. Reducing \eqref{eq:u1gac} modulo 12 and dividing the resulting equation by 2, we conclude that the expression on the right-hand side of \eqref{qe:u1mixedan} vanishes modulo 6. Thus the mixed anomaly coefficient~$k_{g^2 \SP^2}$ is divisible by 6, and consequently the 2-group structure constant $\hat \kappa_\SP = {k_{g^2 \SP^2} \over 6}$ is an integer.

\subsection{'t Hooft Anomalies in the Presence of 2-Group Global Symmetries}\label{ssec:2gphooft}

In a 6d theory with symmetries $G^{(0)}$ and $U(1)_B^{(1)}$, we can consider the following counterterm for the background fields $A_G^{(1)}$ and $B^{(2)}$ (we will call it a GS counterterm),
\begin{equation}\label{GScounterterm}
S_\text{GS}=in \int B^{(2)}\wedge c_2(F_G)~, \qquad n\in \Z . 
\end{equation}
Here the quantization condition on $n$ stems from the requirement that the GS counterterm be invariant under large $U(1)_B^{(1)}$ background gauge transformations. The choice of the coefficient $n$ amounts to a choice of regularization scheme, and is therefore not intrinsic to the theory under discussion.\footnote{~The situation is more intricate if the theory under discussion is a low-energy effective theory, as is often the case in this paper. Then the coefficient $n_\text{IR}$ with which the counterterm appears in the IR Wilsonian effective action need not be the same as the UV coefficient $n_\text{UV}$. Even though $n_\text{IR}$ inherits the scheme ambiguity of $n_\text{UV}$, their difference $n_\text{UV} - n_\text{IR}$ is a meaningful physical quantity, which arises from integrating out massive degrees of freedom. Moreover, integrating out massless or topological degrees of freedom in the IR may lead to an effective non-integral value for $n_\text{IR}$. See \cite{Closset:2012vg,Closset:2012vp} for a detailed discussion of these points in the context of 3d Chern-Simons terms, and \cite{Cordova:2018cvg} for a closely related discussion of GS terms in 4d.} 

If the symmetries (including Poincar\'e symmetry $\SP^{(0)}$) are tied together in non-trivial 2-group $\big(G^{(0)} \times \SP^{(0)}\big) \times_{\hat \kappa_G, \hat \kappa_\SP} U(1)_B^{(1)}$, then $B^{(2)}$ shifts under $G^{(0)}$ background gauge transformations and local Lorentz transformations as in \eqref{eq:Bshiftsintro}, 
\begin{equation}\label{eq:Bshiftsagain}
B^{(2)} \rightarrow B^{(2)}   +\frac{\hat \kappa _G}{4\pi}\Tr\left( \lambda _G^{(0)}dA_G^{(1)}\right)+\frac{\hat \kappa _\SP}{16\pi} \tr \left(\theta ^{(0)}d\omega ^{(1)}\right)~.
\end{equation}
Substituting into \eqref{GScounterterm}, we see that the GS counterterm shifts as follows,
\begin{equation}\label{GScountertermc}
S_\text{GS} \rightarrow S_\text{GS}  + i n  \int \left(\frac{\hat \kappa _G}{4\pi}\Tr\left( \lambda _G^{(0)}dA_G^{(1)}\right)+\frac{\hat \kappa _\SP}{16\pi} \tr \left(\theta ^{(0)}d\omega ^{(1)}\right)\right) \wedge c_2(F^{(2)}_G)~.
\end{equation}
The variation of the GS counterterm therefore contributes to the anomaly polynomial as follows,
\begin{equation}\label{eq:deltags}
\Delta \CI^{(8)} = n \hat \kappa_G c_2(F_G^{(2)})^2 - {n \hat \kappa_\SP \over 4} p_1(T)c_2(F_G^{(2)})~.
\end{equation}
This is simply the GS mechanism for background fields at work: \eqref{GScounterterm} is the GS term, while the 2-group symmetry furnishes $B^{(2)}$ with GS shifts under background 0-form gauge transformations, resulting in a reducible contribution \eqref{eq:deltags} to the anomaly polynomial.

We can similarly consider a GS counterterm involving $p_1(T)$ rather than $c_2(F_G^{(2)})$, 
\begin{equation}\label{GScountertermii}
S'_\text{GS}={in' \over 4} \int B^{(2)}\wedge p_1(T)~, \qquad n' \in \Z ~.
\end{equation}
As before the quantization of $n'$ follows by demanding invariance of this GS term under large $B^{(2)}$ gauge transformations.\footnote{~Under such a gauge transformation, $B^{(2)}$ shifts by $F_\Lambda^{(2)} = d\Lambda_B^{(1)}$ with $\Lambda_B^{(1)}$ a conventionally normalized $U(1)^{(0)}$ gauge field and $F_\Lambda^{(2)}$ its Maxwell field strength, i.e.~${1 \over 2 \pi} F_\Lambda^{(2)}$ defines an integral cohomology class. To show that ${1 \over 8 \pi} \int F_\Lambda^{(2)} \wedge p_1(T) \in \Z$ on a spin 6-manifold we combine the integrality of ${1 \over 2 \pi}  F_\Lambda^{(2)}$ with the Atiyah-Singer index theorem, which states that the index of a unit-charge Dirac operator is given by
\begin{equation}
\text{index} = \int \bigg( {1 \over 3!} \Big({F_\Lambda^{(2)} \over 2 \pi}\Big)^3 - {1 \over 24}  \Big({F_\Lambda^{(2)} \over 2 \pi}\Big) p_1(T) \bigg) \in \Z~.
\end{equation}
} (A very similar quantization condition was discussed in \cite{Intriligator:2014eaa}, albeit in the context of GS terms for dynamical, self-dual 2-form gauge fields.) Substituting \eqref{eq:Bshiftsagain} into this counterterm leads to another reducible contribution to the anomaly polynomial,
\begin{equation}\label{eq:deltagsprime}
\Delta \CI^{(8)} = {n' \hat \kappa_G \over 4} c_2(F_G^{(2)}) p_1(T) - {n' \hat \kappa_\SP \over 16} p_1(T)^2~. 
\end{equation}
Adding \eqref{eq:deltags} and \eqref{eq:deltagsprime}, we obtain the full form of the terms in the anomaly polynomial that are rendered scheme dependent by the GS counterterms \eqref{GScounterterm} and \eqref{GScountertermii},
\begin{equation}\label{eq:thanomschemedep}
\Delta \CI^{(8)}_\text{total} = n \hat \kappa_G c_2(F_G^{(2)})^2 + {n' \hat \kappa_G - n \hat \kappa_\SP \over 4} c_2(F_G^{(2)}) p_1(T) - {n' \hat \kappa_\SP \over 16} p_1(T)^2~.
\end{equation}
Only the remainder of the anomaly polynomial $\CI^{(8)}$ modulo the counterterm ambiguity $\Delta \CI^{(8)}_\text{total}$ encodes genuine 't Hooft anomalies that are subject to 't Hooft anomaly matching (see \cite{Cordova:2018cvg} for a detailed discussion with examples in 4d).

\subsection{Example of a 6d Sigma Model with Continuous 2-Group Symmetries}

Given a sigma model with target space $X$, the Hodge dual of an integral cohomology class in $H^p(X, \R)$, pulled back to spacetime, gives rise to a conserved $(d-p)$-from current with quantized charges in $d$ spacetime dimensions. In $d = 6$, this means that continuous $U(1)_X^{(1)}$ 1-form symmetries can arise if $X$ has non-trivial $H^4(X, \R)$. An example is the sigma model with target space $X = S^4$. 

To see that the $S^4$ sigma model can be modified so that its $U(1)_X^{(1)}$ symmetry participates in a 2-group, it is convenient to engineer an RG flow from a 6d gauge theory with 2-group symmetry that realizes the $S^4$ sigma model in the IR. This can be done using a dynamical $g^{(0)} = SU(2)$ gauge theory coupled to two complex scalar Higgs fields $H_1, H_2$, each of which is in the doublet representation of $SU(2)$. Prior to gauging the $SU(2)$ symmetry, the Higgs fields comprise eight real scalars transforming in the vector representation of an $SO(8)$ flavor symmetry. We can add an $SO(8)$-invariant potential $V(H^\dagger_1 H_1 + H_2^\dagger H_2)$ that induces a vev $H^\dagger_1 H_1 + H_2^\dagger H_2 = v^2$. Prior to gauging, this vev breaks the $SO(8)$ symmetry to $SO(7)$, leading to an $S^7$ sigma model for the massless NG bosons in the deep IR. The gauged $SU(2)$ symmetry acts on $S^7$ via the Hopf action, and performing the gauge quotient reduces $S^7$ to $S^4$. In the Higgs phase, topologically non-trivial configurations of the $SU(2)$ gauge fields are tied to winding configurations of the scalars $H_i~(i = 1,2)$. (One way to see this is to demand that the action be finite, which requires the gauge-covariant derivatives $D_\mu H_i$ to vanish at infinity.) Consequently, the instanton density $c_2(f_g^{(2)})$ of the $SU(2)$ gauge theory flows to the pullback to spacetime of the $S^4$ volume form. In other words, the continuous $U(1)_X^{(1)}$ 1-form symmetry of the $S^4$ sigma model descends from the instanton 1-form symmetry of the $SU(2)$ gauge theory we started with. 

It is now straightforward to enrich the preceding model so that the $U(1)_X^{(1)}$ symmetry participates in a 2-group. For instance, we could add $N_f$ fundamental Dirac fermions to the $SU(2)$ gauge theory. This leads to a new $G^{(0)} = SU(N_f)^{(0)}_L \times SU(N_f)_R^{(0)}$ flavor symmetry acting on the chiral and anti-chiral parts of the Dirac fermions. Since $G^{(0)}$ has mixed anomalies with the $g^{(0)} = SU(2)$ gauge symmetry, we obtain a 2-group of the form (here we use \eqref{eq:ffkgkp})
\begin{equation}
\left(SU(N_f)^{(0)}_L \times SU(N_f)_R^{(0)} \right) \times_{\hat \kappa_L, \hat \kappa_R} U(1)_X^{(1)}~, \qquad \hat \kappa_L = - \hat \kappa_R = -1~.
\end{equation}

\subsection{2-Group Symmetries and 't Hooft Anomalies on String Worldsheets} 

For completeness we mention that (just as in four dimensions \cite{Cordova:2018cvg}) 2-group symmetry fixes certain 't Hooft anomalies on the 2d world sheets of strings charged under the $U(1)_B^{(1)}$ 1-form symmetry. This can be seen by integrating the 2-group non-conservation equations \eqref{modcons} over the four-dimensional space $\Sigma_\perp$ transverse to the world sheet. Then $\int_{\Sigma_\perp} * J_B^{(2)} = Q$ measures the $U(1)_B^{(1)}$ charge of the string. 

A quick way to obtain the world sheet anomalies in the form of a 4-form anomaly polynomial $\CI^{(4)}_\text{string}$ involves looking at the instanton strings in the gauge theory examples discussed in section \ref{ssec:moreImixed}. If we integrate the anomaly 8-form polynomial $\CI^{(8)}_\text{mixed}$ in \eqref{mixedcoeffs} over $\Sigma_\perp$ we can replace $\int_{\Sigma_\perp} c_2(f_g^{(2)}) \rightarrow Q$. We therefore find that 
\begin{equation}
\CI^{(4)}_\text{string} = Q \left(k_{g^2G^2} c_2(F^{(2)}_G)+\frac{k_{g^2\SP^2}}{24}p_1(T)\right) = Q \left(- \hat \kappa_G c_2(F^{(2)}_G)+{\hat \kappa_\SP \over 4} p_1(T)\right)~.
\end{equation}
In the second equation we have substituted the anomaly coefficients by the 2-group structure constants using \eqref{eq:2gpconstanom}, and this equation is a model-independent relation for the world-sheet 't Hooft anomalies of a charge-$Q$ string in any 6d theory with 2-group symmetry.

Note that 6d SCFTs also have strings, but they are not charged under a global $U(1)_B^{(1)}$ symmetry, given that such a symmetry does not exist in these theories (see section \ref{ssec:no1fmscftint}). Rather, the strings in 6d SCFTs carry gauge charge under dynamical 2-form gauge fields, and the GS terms for these gauge fields determine the 't Hooft anomalies on the string world sheet (see for instance~\cite{Shimizu:2016lbw,DelZotto:2018tcj} and references therein). As explained in section \ref{ssec:gsintro}, and following \cite{Ohmori:2014kda}, these GS terms must be chosen to cancel any apparent 2-group symmetry of the low-energy gauge theory.

 \section{6d Little String Theories and 2-Group Global Symmetries}\label{sec:LSTs}
 
In this section we examine the SUSY gauge theories that describe the deep IR of 6d LSTs (focusing on the examples mentioned in section \ref{ssec:introlst}) and show that they possess continuous 2-group global symmetries arising from mixed gauge-global anomalies, as in sections~\ref{ssec:mixanom2gpintro} and~\ref{ssec:moreImixed}. We do not know the extent to (and the precise sense in) which these 2-group symmetries may extend to the full LSTs. At worst, they may only be accidental symmetries that emerge in the deep IR, but are not present microscopically. The basic challenge is that 2-group symmetry is not immediately visible at the level of the symmetry charge algebra. In fact, all symmetries that we will discuss in the context of LSTs are believed to be global symmetries at the level of global charges. This holds for the (super-) Poincar\'e symmetries, the flavor and $R$-symmetries, and even for the instanton 1-form symmetry $U(1)_B^{(1)}$, whose charge can act as a higher-form version of a central charge in the SUSY algebra, where it leads to a BPS bound for instanton strings (see \cite{Dumitrescu:2011iu} for an in-depth discussion). 

By contrast, continuous 2-group symmetries are based on the local notion of current algebra. On one hand, it is not clear to what extent this notion retains meaning in the fundamentally non-local LSTs. On the other hand, the 2-group Ward identities (as explained in~\cite{Cordova:2018cvg}, these follow from the 2-group non-conservation equations in \eqref{modcons}) can be Fourier transformed to yield well-defined relations between momentum-space Green functions, which are also meaningful in LSTs (see for instance \cite{Aharony:1999ks, Kapustin:1999ci}). We view it as an interesting problem to elucidate the status of the 2-group global symmetries we identify below (using the low-energy gauge theory description) for the full LSTs. (See \cite{delzohmToApp} for some upcoming work in this direction.) Similar caveats apply to the notion of 't Hooft anomalies in LSTs (which we will also analyze using the low-energy gauge theory description). 
 
\subsection{Anomalies and 2-Group Structure Constants in Supersymmetric Theories} 

In this section we summarize the anomaly polynomials (see appendix A of~\cite{Ohmori:2014kda} for a useful summary), and compute the 2-group structure constants, of weakly-coupled 6d $\CN=(1,0)$ theories containing hyper, vector, and tensor multiplets. We do not consider the contribution of possible GS terms for tensor multiplets, which will be absent in the LST discussed in this section, but will play an important role in our discussion of SCFTs in section \ref{sec:scfts}. 

The theories we consider have a compact simple dynamical gauge group $g^{(0)}$, and hence an associated $U(1)_B^{(1)}$ instanton 1-form symmetry. In addition to Poincar\'e symmetry $\SP^{(0)}$ and its supersymmetric extension, all theories also have an $SU(2)_R$ symmetry under which the supercharges transform in the fundamental $\square$ representation. In addition, they can also have a non-$R$ flavor symmetry $G^{(0)}$, which commutes with the supercharge. 

Hyper multiplets contain chiral fermions, which can transform in a representation $\rho_g \otimes \rho_G$ of the gauge and non-$R$ flavor symmetry, but are neutral under the $SU(2)_R$ symmetry. The anomaly polynomial of such a hyper multiplet is thus given by \eqref{sixdanomg}, which can be simplified using \eqref{eq:casimir},
\begin{equation}\label{hyperanom}
\begin{split}
\CI^{(8)}_\text{hyper} (\rho_g, \rho_G) =~& \frac{|\rho _G|}{4! (2\pi)^4}\Tr _{\rho _g} (f_g^{(2)})^4+\frac{|\rho _g|}{4!(2\pi)^4}\Tr _{\rho _G} (F_G^{(2)})^4 + T_2(\rho_g) T_2(\rho_G) c_2(f_g^{(2)}) c_2(F_G^{(2)}) \\
 &+\frac{1}{24} \, p_1(T) \, \left(|\rho _g| T_2(\rho_G) c_2(F_G^{(2)}) + |\rho _G| T_2(\rho_g) c_2(f_g^{(2)}) \right) \\
 & +\frac{|\rho _G||\rho _g|}{2^36!} \Big(7p_1(T)^2-4p_2(T)\Big).
 \end{split}
\end{equation}

Vector multiplets contain anti-chiral fermions (i.e.~gauginos) that transform in the adjoint representation of the gauge group $g^{(0)}$ and in the spin-$\half$ fundamental representation $\square$ of $SU(2)_R$. (They are neutral under the non-$R$ flavor symmetry $G^{(0)}$.) Moreover, the gauginos (just like the 6d supercharges) obey a symplectic-Weyl reality condition, which requires us to multiply the anomaly polynomial by $\half$. Bearing in mind that anti-chiral fermions contribute anomalies of opposite sign (as in \eqref{eq:antichii}), that adjoint representation of $g^{(0)}$ has $T_2(\text{adj}) = 2 h^\vee_g$ and dimension $\text{dim}(g^{(0)})$, and the $SU(2)_R$ group theory relation $\Tr_\square ( F_R^{(2)})^4 = \half (\Tr_\square (F_R^{(2)})^2)^2$, we find that the anomaly polynomial of a vector multiplet is 
\begin{equation}\label{sixdvanom}
\begin{split}
\CI _{ \text{vector}}(g^{(0)}) =~& - {1 \over 4! (2\pi)^4} \Tr_\text{adj} (f_g^{(2)})^4 - h^\vee_g \left(c_2(R) +\frac{1}{12} \, p_1(T)\right) c_2(f_g^{(2)}) \\
&- \text{dim}(g^{(0)}) \left(\frac{1}{4!} c_2(R)^2+\frac{1}{48}c_2(R)p_1(T)+ \frac{1}{2^3 6!} (7p_1(T)^2-4p_2(T))\right)~.
\end{split}
\end{equation}

For future reference, we write down the anomaly polynomial for a tensor multiplet. In addition to a chiral fermion transforming in the $\square$ of the $SU(2)_R$ symmetry (but neutral under $g^{(0)}, G^{(0)}$, this also receives a purely gravitational contribution from the chiral 2-form gauge field residing in the tensor multiplet \cite{AlvarezGaume:1983ig}. Altogether (see for instance the second line in Table \ref{known}, or equation (A.4) in~\cite{Ohmori:2014kda}), 
\begin{equation}\label{eq:Itensor}
\CI _\text{tensor}=\frac{1}{4!}c_2(R)^2+\frac{1}{48} c_2(R)p_1(T)+\frac{1}{2^3 6!}\left(23p_1(T)^2-116 p_2(T)\right).
\end{equation}

In order to compute the 2-group structure constants, we need the following mixed anomaly coefficients,
\begin{equation}\label{Imixedagain}
\CI^{(8)}_\text{mixed} = c_2(f_g^{(2)}) \left(k_{g^2 G^2} c_2(F_G^{(2)}) + k_{g^2 R^2} c_2(R) + {k_{g^2 \SP^2} \over 24} p_1(T)\right)~.
\end{equation}
Only the hyper and vector multiplets in \eqref{hyperanom} and \eqref{sixdvanom} contributed to these mixed anomalies.  Summing over all hyper multiplet representations $(\rho_{g, i}, \rho_{G, i})$, we find 
 \begin{equation}\label{eq:susymxanom}
 k_{g^2G^2}=\sum_{\text{hypers }i} T_2(\rho _{g,i})T_2(\rho _{G,i})~, \qquad k_{g^2R^2}=-h_g^\vee~,  \qquad k_{g^2\SP^2}=-2h_g^\vee +\sum_{\text{hypers }i} |\rho _{G,i} |T_2(\rho _{g,i})~.
 \end{equation}
The 2-group structure constants are then given in terms of these mixed anomalies by \eqref{eq:2gpconstanom},
\begin{equation}\label{kappasfree}
\begin{split}
& \hat \kappa _G = - k_{g^2G^2} = -\sum_{\text{hypers }i} T_2(\rho _{g,i})T_2(\rho _{G,i})~, \\
& \hat \kappa _R = - k_{g^2R^2}=h_g^\vee~,\\
& \hat \kappa _\SP = {k_{g^2\SP^2} \over 6} =-{h_g^\vee \over 3} +{1 \over 6} \sum_{\text{hypers }i} |\rho _{G,i} |T_2(\rho _{g,i})~.
\end{split}
\end{equation}
Whenever any of these structure constants are non-zero, the global symmetries form a non-trivial 2-group. As discussed in section \ref{ssec:2gphooft}, this subjects the 't Hooft anomalies $\CI^{(8)}_\text{global}$ to scheme-dependent shifts of the form \eqref{eq:thanomschemedep}, which are proportional to the 2-group structure constants.

\subsection{$\CN =(1,1)$ Little String Theories}

As was briefly reviewed in section \ref{ssec:introlst}, the $\CN=(1,1)$ LSTs flow at low energies to maximally supersymmetric Yang-Mills theory with ADE gauge group $g^{(0)}$. (The anomaly coefficients and 2-group structure constants listed below are valid for any $g^{(0)}$.) In $\CN=(1,0)$ language, these are theories with vector multiplet and a hyper multiplet in the adjoint representation of $g^{(0)}$. Together the chiral and anti-chiral adjoint fermions in these two multiplets form an Dirac fermion in the adjoint of $g^{(0)}$, so that all gauge anomalies automatically vanish. 

The 0-form symmetries are the super-Poincar\'e symmetry, as well as an $SU(2)^{(0)}_L \times SU(2)^{(0)}_R$ symmetry under which the supercharges transform in the bi-fundamental representation $\left(\square, \square\right)$ (i.e.~in the spin $(\half, \half)$ representation). From the $\CN=(1,0)$ point of view, $SU(2)_L^{(0)}$ (abbreviated by ``$L$'') is a flavor symmetry (it commutes with the $\CN=(1,0)$ supercharges), while $SU(2)_R^{(0)}$ (abbreviated ``$R$'') is the $\CN=(1,0)$ $R$-symmetry. The $\CN=(1,0)$ hyper multiplet is a half-hyper-multiplet in the $\square$ of $SU(2)_L$, i.e.~there are four real hyper-multiplet scalars transforming as bi-fundamentals of $SU(2)_L^{(0)} \times SU(2)_R^{(0)}$. Referring to \eqref{hyperanom} (multiplied by $\half$ because we have a half-hyper-multiplet) and \eqref{sixdvanom}, we obtain the mixed and global anomalies of the theory,
\begin{equation}\label{oneoneanom}
\begin{split}
& \CI^{(8)}_\text{mixed} = h^\vee _g  \Big(c_2(L)-c_2(R)\Big)c_2(f_g^{(2)})~,\\
& \CI^{(8)}_\text{global} = \text{dim}(g^{(0)}) \left(\frac{1}{4!} \left(c_2(L)^2-c_2(R)^2\right)+\frac{1}{48}\left(c_2(L)-c_2(R)\right)p_1(T)\right)~.
\end{split}
\end{equation}
Comparing with \eqref{kappasfree}, we find the following 2-group structure constants, 
\begin{equation}\label{oneonetwogroup}
\hat \kappa _{L}=-\hat \kappa _R=-h^\vee _g~, \qquad \hat \kappa _\SP =0.
\end{equation}
Thus the $SU(2)_L^{(0)} \times SU(2)_R^{(0)}$ symmetry participates in a 2-group with the $U(1)_B^{(1)}$ instanton 1-form symmetry, but the Poincar\'e symmetry does not. As we briefly recalled below \eqref{kappasfree} above (see section \ref{ssec:2gphooft} for more detail) this truncates those 't Hooft anomalies in $\CI^{(8)}_\text{global}$ that can be (partially) absorbed using GS counterterms involving the background field $B^{(2)}$.

\subsection{The $\CN=(1,0)$ Little String Theory of Small $SO(32)$ Instantons}

As was shown in 
\cite{Witten:1995gx, Seiberg:1997zk}, the low-energy gauge theory that describes the deep IR of the 6d LST obtained by decoupling gravity for $N$ coincident small heterotic $SO(32)$ instantons is an~$\CN=(1,0)$ gauge theory with gauge group $g^{(0)} = Sp(N)$, a global $SU(2)_R^{(0)}$ symmetry under which the supercharges transform in the fundamental $\square$ representation, and a $G^{(0)} = SU(2)_L^{(0)} \times SO(32)^{(0)}$ flavor symmetry that commutes with the supercharges. The matter consists of a half-hyper-multiplet in the bi-fundamental representation $(\square, \square)$ of $Sp(N) \times SO(32)$, and another half-hyper-multiplet in the irreducible two-index antisymmetric tensor representation of the $Sp(N)$ gauge group and the $\square$ representation of $SU(2)_L$. 

It was shown in~\cite{Schwarz:1995zw} that this matter content cancels all gauge anomalies. The mixed anomalies can be computed by substituting the above matter content into \eqref{eq:susymxanom}, taking care to multiply by $\half$ for the half-hyper-multiplets,\footnote{~Here we need the fact that $T_2 = 2(N-1)$ for the irreducible two-index antisymmetric tensor representation of $Sp(N)$. Note the simple special cases $N=1$, where this representation is a singlet with $T_2 = 0$, and $N=2$, where this representation is equivalent to the fundamental $SO(5)$ vector representation, with $T_2 = 2$.}
 \begin{equation}
 k_{g^2L^2}=N-1~, \qquad k_{g^2R^2}=-h_g^\vee = - (N+1)~,  \qquad k_{g^2 (SO(32))^2}=1~, \qquad k_{g^2\SP^2}=12~.
 \end{equation}
This leads to the following 2-group structure constants,
\begin{equation}
\begin{split}
& \hat \kappa _L=- k_{g^2L^2} = -(N-1)~,   \qquad\;\;\;\; \hat \kappa _R= -k_{g^2R^2}=h_g^\vee~, \\
& \hat \kappa _{SO(32)} = -k_{g^2 (SO(32))^2}=-1~, \qquad \hat \kappa _\SP={k_{g^2\SP^2} \over 6}=2~,
\end{split}
\end{equation}
all of which are integers. In the case of $\hat \kappa_\SP$, this is non-trivial, but ultimately guaranteed by the fact that the matter content cancels all pure $Sp(N)$ gauge anomalies (see section \ref{ssec:quantcond}). We see that all possible 2-group structure constants are generically non-zero (for $N \geq 2$), so that all continuous 0-form symmetries (including the Poincar\'e symmetry) participate in a non-trivial 2-group with the instanton 1-form symmetry $U(1)_B^{(1)}$.  As in the discussion below~\eqref{oneonetwogroup}, this has the effect of truncating certain 't Hooft anomalies of the theory (see also section \ref{ssec:2gphooft}).

\section{Positivity of the $a$-Type Weyl Anomaly in 6d SCFTs}\label{sec:scfts}

In this section we argue that a unitary $\CN=(1,0)$ SCFT with a tensor branch has positive $a$-type Weyl anomaly, $a_\text{SCFT} > 0$. As in section \ref{ssec:aposintro} we first focus on the rank-1 case, before discussing the generalization to higher rank. 

\subsection{Proof For Rank-1 Theories}\label{ssec:rk1pf}

Recall from section \ref{ssec:aposintro} that the $a$-type Weyl anomaly $a_\text{SCFT}$ of a 6d $\CN=(1,0)$ SCFT with a rank-1 tensor branch (described in the deep IR by $n_T = 1$ tensor multiplets, $n_V = \text{dim}(g^{(0)})$ vector multiplets in the adjoint representation of a compact simple gauge group $g^{(0)}$, and hyper multiplet matter) is given by \eqref{eq:ascft} and \eqref{eq:deltagsintro}, which we repeat here,
\begin{equation}\label{eq:ascftagain}
\begin{split}
& a_\text{SCFT} = a_\text{IR} + \Delta a~,\\
& a_\text{IR} = \frac{11}{210}n_H+\frac{199}{210}n_T-\frac{251}{210}n_V~,\\
&\Delta a =  -\frac{192}{7 k_{(g^2)^2}} \left(k_{g^2 R^2} - {k_{g^2 \SP^2} \over 24}  \right)^2 > 0~.
\end{split}
\end{equation} 
The fact that $\Delta a$ can be computed in this way non-trivially follows from the absence of continuous 2-group symmetries in 6d SCFTs (see sections \ref{ssec:no1fmscftint} and \ref{ssec:aposintro} for details and references). 

We would like to show that $a_\text{SCFT}$ is positive, i.e.~that $\Delta a > 0$ is always sufficiently large to overpower any potentially negative contribution to $a_\text{IR}$ coming from vector multiplets. We will show that this follows from general properties of the low-energy theory on the tensor branch. To this end, we must compute the 1-loop anomaly coefficients $k_{(g^2)^2}$, $k_{g^2 R^2}$, and $k_{g^2 \SP^2}$ due to massless vector and hyper multiplets on the tensor branch that appear in \eqref{eq:ascftagain}, normalized as in \eqref{eq:6d1lgauge} and \eqref{eq:i1lmixed}, 
\begin{equation}\label{eq:1lptb}
\CI^{(8)}_\text{1-loop} \supset \half \, k_{(g^2)^2} \, c_2(f_g^{(2)})^2 + k_{g^2 R^2} \, c_2(f_g^{(2)}) \, c_2(R) + {k_{g^2 \SP^2} \over 24} \, c_2(f_g^{(2)}) \, p_1(T)~.
\end{equation}

We have already computed the anomaly coefficients $k_{g^2 R^2}$ and $k_{g^2 \SP^2}$ in \eqref{eq:susymxanom},\footnote{~Here we ignore the flavor symmetry $G^{(0)}$ and take $\rho_G$ to be the trivial representation, but still sum over all hyper multiplets. In other words, if IR gauge theory does in fact have a $G^{(0)}$ symmetry, we treat a hyper multiplet transforming in the $\rho_G$ representation as $|\rho_G|$ singlets and sum over all of them.}
 \begin{equation}\label{eq:freekmixedtb}
k_{g^2R^2}=-h_g^\vee~,  \qquad k_{g^2\SP^2}=-2h_g^\vee +\sum_{\text{hypers }i} T_2(\rho _{g,i})~.
 \end{equation}
Since $T_2(\rho) \geq 0$, we obtain the following lower bound for the quadratic expression that appears in the formula~\eqref{eq:ascftagain} for $\Delta a$, \begin{equation}\label{eq:bound1}
\left(k_{g^2 R^2} - {k_{g^2 \SP^2} \over 24}  \right)^2 \geq \left({11 h_g^\vee \over 12}\right)^2~.
\end{equation}
Note that this lower bound only depends on the gauge group $g^{(0)}$, but not on the hyper multiplet matter content. The value of $h_g^\vee$ is tabulated in table \ref{adjoint} for all compact simple gauge groups.

\bigskip

\begin{table}[h]
\centering
\begin{tabular}{!{\VRule[1pt]}c|!{\VRule[1pt]}c!{\VRule[1pt]}c!{\VRule[1pt]}c!{\VRule[1pt]} c!{\VRule[1pt]}c!{\VRule[1pt]}c!{\VRule[1pt]}c!{\VRule[1pt]}c!{\VRule[1pt]}c!{\VRule[1pt]}c!{\VRule[1pt]}}
\specialrule{1.2pt}{0pt}{0pt}
{$g^{(0)}$} & $SU(N \geq 4)$ &  $SO(N \geq 5)$ & $Sp(N)$ & $SU(2)$& $SU(3)$& $G_2$& $F_4$& $E_6$& $E_7$& $E_8$  \\
\specialrule{1.2pt}{0pt}{0pt}
\specialrule{1.2pt}{0pt}{0pt}
{$\text{dim}(g^{(0)})$} & $N^2-1$ &  ${N(N-1) \over 2}$ & $N (2N+1)$ & $3$& $8$& $14$& $52$& $78$& $133$& $248$  \\
\specialrule{1.2pt}{0pt}{0pt}
{$h_g^{\vee}$} & $N$ &  $N-2$ & $N+1$ & $2$& $3$& $4$& $9$& $12$& $18$& $30$  \\
\specialrule{1.2pt}{0pt}{0pt}
{$u_g$} & $2$ &  $4$ & $1$ & $8 \over 3$& $3$& $10 \over 3$& $5$& $6$& $8$& $12$  \\
\specialrule{1.2pt}{0pt}{0pt}
\end{tabular}
\caption{Dimension $\text{dim}(g^{(0)})$, dual Coxeter number $h_g^\vee$, and $u_g$ coefficient (defined in \eqref{eq:ugdef}) for all compact simple gauge groups $g^{(0)}$. The groups $SU(2)$ and $SU(3)$ require special treatment because they do not have quartic Casimirs.}
\label{adjoint}
\end{table}

\bigskip

We must now extract the reducible gauge-anomaly coefficient $k_{(g^2)^2}$ in \eqref{eq:1lptb} from the anomaly polynomials \eqref{hyperanom} and \eqref{sixdvanom} for hyper and vector multiplets,
\begin{equation}\label{eq:hypvecmixg}
\CI^{(8)}_\text{gauge, hyper+vector}  = {1 \over 4! (2 \pi)^4}  \sum_{\text{hypers }i} \Tr_{\rho_g, i} (f_g^{(2)})^4 - {1 \over 4! (2 \pi)^4} \Tr_\text{adj} (f_g^{(2)})^4~.
\end{equation}
Recall from \eqref{eq:mixedkneg} that the reducible gauge anomaly coefficient $k_{(g^2)^2}$ must be negative in 6d SCFTs, $k_{(g^2)^2}  < 0$, since only such a negative anomaly can be canceled using a field-theoretic GS mechanism with a $\CN=(1,0)$ tensor multiplet. It follows from \eqref{eq:hypvecmixg} that hyper multiplets contribute positively to the gauge anomaly, and hence to $k_{(g^2)^2}$, while vector multiplets contribute negatively.\footnote{~This bounds the overall amount of charged hyper-multiplet matter that is possible for a given gauge group, as originally pointed out in \cite{Seiberg:1996qx}.} This leads to the bounds
\begin{equation}\label{eq:bound2}
k^\text{vector}_{(g^2)^2} \leq k_{(g^2)^2} < 0~, 
\end{equation}
where $k^\text{vector}_{(g^2)^2}$ is the vector-multiplet contribution to the reducible gauge anomaly coefficient.
 
To compute $k^\text{vector}_{(g^2)^2}$, we must expand $\Tr_\text{adj} (f_g^{(2)})^4$ in terms of Chern classes and extract the reducible term $\sim c_2(f_g^{(2)})^2$, which involves a group-theoretic factor. Following appendix~A of \cite{Ohmori:2014kda}, we define a constant $u_g >0$ (which only depends on the gauge group $g^{(0)}$) via
\begin{equation}\label{eq:ugdef}
\Tr_\text{adj} (f_g^{(2)})^4 \bigg|_\text{reducible} \equiv 3 u_g (\Tr (f_g^{(2)})^2)^2~.
\end{equation}
The coefficient $u_g$ is listed in table \ref{adjoint} for all compact simple gauge groups. Since $c_2(f_g^{(2)}) = {1 \over 8 \pi^2} \Tr(f_g^{(2)})^2$ (see~\eqref{instcurr}), we find that 
\begin{equation}\label{eq:kisu}
k^\text{vector}_{(g^2)^2} = - u_g~.
\end{equation}

We can now put everything together: substituting \eqref{eq:kisu} into the bound \eqref{eq:bound2}, and combining the latter with the previously obtained bound in \eqref{eq:bound1}, we deduce a lower bound for $\Delta a$ in \eqref{eq:ascftagain} that only depends on the gauge group $g^{(0)}$,
\begin{equation}\label{eq:finalbound}
\Delta a \geq {192 \over 7 u_g} \left(11 h_g^\vee \over 12\right)^2 \approx  {23.05 \, (h_g^\vee)^2 \over u_g}~.
\end{equation}
Omitting the manifestly positive hyper and tensor multiplet contributions to $a_\text{IR}$ in \eqref{eq:ascftagain}, and using $n_V = \text{dim}(g^{(0)})$, we also obtain a (crude but sufficient) lower bound on $a_\text{IR}$,
\begin{equation}
a_\text{IR} > -{251 \over 210} \text{dim} (g^{(0)}) \approx - 1.20~  \text{dim}(g^{(0)})~.
\end{equation}
In order to show that $a_\text{SCFT}$ is positive, we must therefore verify that 
\begin{equation}\label{eq:liealgbound}
{23.05 \, (h_g^\vee)^2 \over u_g} > 1.20~  \text{dim}(g^{(0)})~,
\end{equation}
or equivalently that
\begin{equation}
{u_g \text{dim}(g^{(0)}) \over (h_g^\vee)^2} < 19.21~.
\end{equation}
This is comfortably true for all gauge groups listed in table \ref{adjoint} and completes the proof that $a_\text{SCFT} > 0$ for rank-1 theories.

\subsection{Generalization to Higher Rank}

We now generalize the proof that $a_\text{SCFT}> 0$ for unitary $\CN=(1,0)$ SCFTs of rank $r = 1$ in section~\ref{ssec:rk1pf} above to SCFTs of rank $r \geq 2$. We follow the same general approach, but the argument is significantly more involved and requires new ingredients. In particular, we will use some results from \cite{Cordova:2019wns}.

As a first step, we activate all $r$ tensor-multiplet vevs $\phi_i~(i = 1, \ldots, r)$ and assume that the resulting low-energy theory on the tensor branch is described by weakly-coupled tensor, vector, and hyper multiplets.\footnote{~This amounts to the assumption that there are no interacting rank-0 theories. This assumption is motivated by the fact that the only known $\CN=(1,0)$ SCFTs without a tensor branch are theories of free hyper multiplets.} The vector multiplets transform in the adjoint representation of a semi-simple Yang-Mills gauge algebra $\oplus_{a = 1}^{\ell} \, \mathfrak{g}_a$, where each $\mathfrak{g}_a$ is a compact simple Lie algebra.\footnote{~Recall that gauge-anomaly cancellation considerations and supersymmetry imply that abelian gauge fields do not occur on the tensor branch of 6d SCFTs.} Here the number~$\ell$ of simple gauge algebras $\mathfrak{g}_a$ must be less than the number $r$ of tensor multiplets, i.e.~$\ell \leq r$, because every such gauge algebra must be paired with at least one tensor multiplet. The reasons are entirely analogous to those explained for the rank-1 case in section~\ref{ssec:gsintro}.

Flux quantization for the self-dual 3-form field-strengths $h_i^{(3)} = *h^{(3)}$ of the 2-form gauge fields $b^{(2)}_i$ residing in the same tensor multiplets as the $\phi_i$ defines natural integral bases for the tensor multiplets.\footnote{~An integral basis is not unique, since a transformation in $O(r, \Z)$ preserves the integrality of the basis.} In such an integral basis, the Dirac pairing $\Omega_{ij} = \Omega_{(ij)}$ for the tensor  multiplets is a real, symmetric matrix with integer entries. Moreover, $\Omega_{ij}$ defines a positive-definite bilinear form, since all the~$b^{(2)}_i$ are chiral 2-form gauge fields with self-dual 3-form field-strengths $h_i^{(3)}$. In such a basis, the GS modification of the Bianchi identities $dh_i^{(3)}$ takes the form
\begin{equation}
dh_i^{(3)} = \CI^{(4)}_i~,
\end{equation}
and the GS contribution $\CI^{(8)}_\text{GS}$ to the anomaly polynomial is given by the following generalization of \eqref{eq:i8gs},
\begin{equation}\label{eq:i8gshrk}
\CI_\text{GS}^{(8)} = \half \, \Omega_{ij} \, \CI^{(4)}_i \wedge \CI^{(4)}_j~. 
\end{equation}
In general, the $\CI_i^{(4)}$ take the form
\begin{equation}\label{eq:i4apphrk}
\CI_i^{(4)} = \sum_{a = 1}^\ell K_{ia} c_2(f_a^{(2)}) + (\text{background gauge fields})~.
\end{equation}
Here~$f_a^{(2)}$ is the Yang-Mills field strength for the gauge algebra $\mathfrak{g}_a$, while the the coefficients $K_{ia}$ determine the Yang-Mills and GS terms, i.e.~the generalizations of \eqref{eq:introgsterm} and \eqref{eq:introymterm},
\begin{equation}\label{eq:hrkymgs}
S_\text{YM+GS} \sim K_{ia} \int \left( \phi_i \Tr(f_a^{(2)} \wedge * f_a^{(2)}) + i b_i^{(2)} \wedge c_2(f_a^{(2)}) \right)~.
\end{equation}
Here the requirement that every simple gauge algebra $\mathfrak{g}_a$ be paired with at least one tensor amounts to the statement that the $r \times \ell$ matrix $K_{ia}$ has maximal rank $\ell$. Moreover, 
invariance of the GS terms in \eqref{eq:hrkymgs} under large gauge transformations of the $b_i^{(2)}$ imposes quantization conditions on the coefficients $K_{ia}$.

We now argue that we can reduce our problem to the case where $r = \ell$ and $K_{ia}$ is a square matrix of maximal rank (i.e.~it is an invertible matrix). To see this, note that in general, there are $r-\ell$ tensor multiplets that are not paired with Yang-Mills gauge fields. If we only activate tensor-multiplet vevs lying in this $(r-\ell)$-dimensional subspace of the full tensor branch, while setting the remaining $\ell$ tensor-multiplet vevs to zero, then the only low-energy degrees of freedom are weakly-coupled tensor and hyper multiplets, as well as an interacting rank-$\ell$ SCFT at the origin -- but no vector multiplets. In this case we can apply the $a$-theorem (together with the fact that tensor and hyper multiplets have positive $a$-anomalies) to conclude that the rank-$r$ SCFT in the UV has positive $a$-anomaly if the rank-$\ell$ SCFT at the origin in the IR has positive $a$-anomaly. 

We therefore proceed to prove that $a_\text{SCFT} > 0$ for rank-$r$ SCFTs with $\ell = r$ simple gauge algebras $\mathfrak{g}_a$ on their tensor branch, so that every gauge algebra is paired with precisely one tensor multiplet. In particular, the square matrix $K_{ia}$ (with $i = 1, \ldots, r$ and $a = 1, \ldots, \ell = r$) appearing in \eqref{eq:i4apphrk} and \eqref{eq:hrkymgs} must be invertible. We can now simplify further by performing a change of basis on the tensor-multiplet fields,
\begin{equation}\label{eq:newbasis}
\hat \phi_a = \sum_{i = 1}^r K_{ia} \phi_i~, \qquad \hat b_a^{(2)} = \sum_{i = 1}^r K_{ia} b_i^{(2)}~,
\end{equation}
and similarly for their fermionic superpartners. Note that this change of basis potentially obscures the flux-quantization properties of the 2-form gauge fields, but this will not affect our argument below. In the new basis \eqref{eq:newbasis}, the Yang-Mills and GS terms in \eqref{eq:hrkymgs} take the following simple, diagonal form,
\begin{equation}
S_\text{YM+GS} \sim \sum_{a = 1}^r \int \left( \hat \phi_a \Tr(f_a^{(2)} \wedge * f_a^{(2)}) + i \hat b_a^{(2)} \wedge c_2(f_a^{(2)})\right)~,
\end{equation}
while the Dirac pairing $\hat \Omega_{ab}$ of the $\hat b_a^{(2)}$ is given by
\begin{equation}
\hat \Omega_{ab} = \sum_{i, j = 1}^r \Omega_{ij} K_{ia} K_{jb}~.
\end{equation}
Comparing with \eqref{eq:i8gshrk} and \eqref{eq:i4apphrk}, we see that $\hat \Omega_{ab}$ is nothing but the matrix of reducible gauge anomalies contributed by the GS terms, 
\begin{equation}
\CI^{(8)}_\text{GS} = \half \hat \Omega_{ab} c_2(f_a^{(2)}) \wedge c_2(f_b^{(2)})~.
\end{equation}
Since this GS contribution must cancel the reducible 1-loop gauge anomalies contributed by vector and hyper multiplets on the tensor branch, we conclude that
\begin{equation}\label{eq:1lgaugehrk}
\CI^{(8)}_\text{1-loop, gauge} = (\text{irreducible}) + \half (k_{(g^2)^2})_{ab} c_2(f_a^{(2)}) \wedge c_2(f_b^{(2)})~, \qquad (k_{(g^2)^2})_{ab} = - \hat \Omega_{ab}~.
\end{equation}
Since the Dirac pairing $\hat \Omega_{ab}$ is a positive-definite bilinear form, it follows that $(k_{(g^2)^2})_{ab}$ must be negative definite. (Recall the analogous discussion around \eqref{eq:mixedkneg} for the rank-1 case.) Note that consistency of the low-energy theory demands that the irreducible gauge anomalies indicated in \eqref{eq:1lgaugehrk}, which are of the form $(k_{g^4})_a\, c_4(f_a^{(2)})$, must vanish in their own right. 

As explained in section \ref{ssec:gsintro}, we can fix the background-field dependence of the $\CI_i^{(4)}$ in~\eqref{eq:i4apphrk}, and hence the GS contribution to the 't Hooft anomalies of the SCFT, by demanding the absence of mixed gauge-global anomalies (and thus the absence of continuous 2-group symmetries). We will do this explicitly for the $SU(2)_R$ and gravitational 't Hooft anomalies, which are needed to compute $a_\text{SCFT}$. The relevant gauge-$R$ and gauge-gravity anomaly coefficients are defined as follows,
\begin{equation}
\CI^{(8)}_\text{1-loop, mixed} = (k_{g^2 R^2})_a \, c_2(f_a^{(2)}) \, c_2(R) + {(k_{g^2 \SP^2})_a \over 24} \, c_2(f_a^{(2)})\,  p_1(T)~.
\end{equation}
We thus arrive at the following generalization of \eqref{eq:fullgloabli8} (with $G^{(0)}$ replaced by $SU(2)_R^{(0)}$) to the higher-rank case,
\begin{equation}
\begin{split}
& \CI^{(8)} = \CI^{(8)}_\text{1-loop, global} \\
& - \half \big(k_{(g^2)^2}\big)_{ab}^{-1} \left((k_{g^2 R^2})_a \, c_2(R) + {(k_{g^2 \SP^2})_a \over 24} \,  p_1(T)  \right) \left((k_{g^2 R^2})_b \, c_2(R) + {(k_{g^2 \SP^2})_b \over 24} \,  p_1(T)  \right)~.
\end{split}
\end{equation}
Here $ \big(k_{(g^2)^2}\big)_{ab}^{-1}$ is the matrix inverse of  $\big(k_{(g^2)^2}\big)_{ab}$. Note that $\big(k_{(g^2)^2}\big)_{ab}$ is invertible because its eigenvalues are strictly negative. Together with \eqref{ais} and \eqref{anomalyI}, this leads to the following generalization of \eqref{eq:ascftagain} to the higher-rank case,
\begin{equation}\label{eq:ascfthrk}
\begin{split}
& a_\text{SCFT} = a_\text{IR} + \Delta a~,\\
& a_\text{IR} = \frac{11}{210}n_H+\frac{199}{210}n_T-\frac{251}{210}n_V~,\\
&\Delta a =  -\frac{192}{7} \big(k_{(g^2)^2}\big)_{ab}^{-1} X_a X_b > 0~, \qquad X_a = {(k_{g^2 \SP^2})_a \over 24} - (k_{g^2 R^2})_a~.
\end{split}
\end{equation}

As for the rank-1 case analyzed in section \ref{ssec:rk1pf}, showing that $a_\text{SCFT} > 0$ amounts to proving that the explicit, positive expression for $\Delta a$ in \eqref{eq:ascfthrk} is always sufficiently large to overwhelm the negative contribution to $a_\text{IR}$ that arises from the
\begin{equation}\label{eq:numvechrk}
n_V = \sum_{a = 1}^r \dim(\mathfrak{g}_a)
\end{equation}
vector multiplets on the tensor branch. We begin by proving that each $X_a$ that appears in the formula \eqref{eq:ascfthrk} for $\Delta a$ satisfies the following bound,
\begin{equation}\label{eq:Xbound}
X_a = {(k_{g^2 \SP^2})_a \over 24} - (k_{g^2 R^2})_a \geq {11 h^\vee_a \over 12} > 0~,
\end{equation}
where $h_a^\vee$ is the dual Coxeter number of $\mathfrak{g}_a$. This is analogous to the bound \eqref{eq:bound1} obtained in the rank-1 case.

To prove the lower bound \eqref{eq:Xbound} for $X_a$, we collapse all $r-1$ tensor-multiplet vevs except $\hat \phi_a$ to the origin, i.e.~we study a one-dimensional tensor sub-branch on which only $\hat \phi_a$ is activated.  The low-energy degrees of freedom on this $\hat \phi_a$-branch are a single tensor multiplet (containing $\hat \phi_a$), vector multiplets transforming in the adjoint representation of~$\mathfrak{g}_a$, hyper multiplet matter, and an interacting rank-$(r-1)$ SCFT $\CT_a$ at the origin of the tensor branch. In general, the $\mathfrak{g}_a$ gauge fields couple to both the weakly-coupled vector and hyper multiplets and to the SCFT $\CT_a$, i.e.~they weakly gauge a $\mathfrak{g}_a$ subgroup of the (non-$R$) flavor symmetry of~$\CT_a$. Consequently, anomalies involving the gauge-algebra $\mathfrak{g}_a$ can receive contributions from any of these sources. We can therefore express the 1-loop gauge-$R$ and gauge-gravity anomalies contributed by the low-energy degrees of freedom as follows (this formula generalizes~\eqref{eq:freekmixedtb}), 
 \begin{equation}\label{eq:freekmixedtbhrk}
(k_{g^2R^2})_a=-h_a^\vee + k_{g^2R^2}(\CT_a)~,  \qquad (k_{g^2\SP^2})_a=-2h_a^\vee +\sum_{\text{hypers }i} T_2(\rho_{a,i}) + k_{g^2\SP^2}(\CT_a)~.
 \end{equation}
 Here $k_{g^2R^2}(\CT_a)$ and $k_{g^2\SP^2}(\CT_a)$ are the mixed $\mathfrak{g}_a$-$R$ and $\mathfrak{g}_a$-gravity anomaly coefficients contributed by the SCFT $\CT_a$, while the $\rho_{a, i}$ denote the $\mathfrak{g}_a$ representations of the weakly-coupled hyper multiplets. 

Viewed intrinsically from the point of view of the SCFT $\CT_a$, the anomaly coefficients $k_{g^2R^2}(\CT_a)$ and $k_{g^2\SP^2}(\CT_a)$ are mixed flavor-$R$ and flavor-gravity 't Hooft anomalies, which are converted into mixed gauge-global anomalies once the $\mathfrak{g}_a$ vector multiplets on the tensor branch weakly gauge a $\mathfrak{g}_a$ subgroup of $\CT_a$'s flavor symmetry.\footnote{~Note that one can make the gauging arbitrarily weak by giving the tensor-multiplet scalar $\hat \phi_a$ an arbitrarily large vev. This does not affect the anomaly coefficients we need.} In order to derive the lower bound \eqref{eq:Xbound} for $X_a$, we compute, using \eqref{eq:freekmixedtbhrk}, 
\begin{equation}\label{eq:stuff}
\begin{split}
X_a & = {(k_{g^2 \SP^2})_a \over 24} - (k_{g^2 R^2})_a\\
& = {11 h_a^\vee \over 12} + \left\{{1 \over 24} \sum_{\text{hypers }i} T_2(\rho _{a,i}) + \left({1 \over 24} k_{g^2\SP^2}(\CT_{a}) - k_{g^2R^2}(\CT_{a})\right)\right\}~.
\end{split}
\end{equation}
The bound \eqref{eq:Xbound} now follows from the fact that the expression $\{\cdots\}$ on the right-hand side of \eqref{eq:stuff} is non-negative. For the hyper-multiplet terms this immediately follows from  the fact that $T_2(\rho) \geq 0$. We now explain why the contribution from the interacting SCFT $\CT_{a}$ is also non-negative. 

In \cite{Cordova:2019wns} we showed that in any $\CN=(1,0)$ SCFT $\CT_a$ with a non-$R$ flavor symmetry $\mathfrak{g}_a$, the corresponding flavor current $j_a^{(1)}$ has a 2-point function $\langle j_a^{(1)}(x) j_a^{(1)}(0) \rangle \sim {\tau_a \over x^{10}}$, whose coefficient $\tau_a$ can be expressed in terms of the mixed 't Hooft anomalies of the $\mathfrak{g}_a$ flavor symmetry with the $SU(2)_R$ symmetry and gravity,
\begin{equation}
\tau_a = (\text{positive number}) \left({1 \over 24} k_{g^2\SP^2}(\CT_{a}) - k_{g^2R^2}(\CT_{a})\right)~.
\end{equation}
Here the positive prefactor depends on the precise normalization of $\tau_a$ and will not be important for us. (It can be found in \cite{Cordova:2019wns}.) Since the 2-point function coefficient $\tau_a$ is necessarily non-negative in a unitary SCFT, $\tau_a \geq 0$, it follows that
\begin{equation}\label{eq:posrkrm1}
{1 \over 24} k_{g^2\SP^2}(\CT_{a}) - k_{g^2R^2}(\CT_{a}) \geq 0~.
\end{equation}
Upon weakly gauging the $\mathfrak{g}_a$ flavor-symmetry of the SCFT $\CT_{a}$ using the $\mathfrak{g}_a$ vector multiplets on the one-dimensional tensor sub-branch parametrized by $\hat \phi_a$, it follows from \eqref{eq:posrkrm1} (together with $T_2(\rho) \geq 0$) that the expression $\{\cdots\}$ on the right-hand side of \eqref{eq:stuff} is non-negative. This proves the lower bound for $X_a$ in \eqref{eq:Xbound}, and it concludes our discussion of the $\hat \phi_a$-branch and the rank-$(r-1)$ SCFT $\CT_a$ residing at its origin. We now return to the full $r$-dimensional tensor branch of the rank-$r$ SCFT for which we are trying to prove that $a_\text{SCFT} > 0$. 

In order to obtain a useful lower bound for $\Delta a$, we must control the matrix $\big(k_{(g^2)^2}\big)_{ab}^{-1} $ that appears in~\eqref{eq:ascfthrk}. Recall from \eqref{eq:1lgaugehrk} that $\big(k_{(g^2)^2}\big)_{ab}$ is the matrix of reducible 1-loop gauge anomalies contributed by the weakly-coupled vector and hyper multiplets on the rank-$r$ tensor branch of the SCFT under consideration. It can therefore be computed explicitly from the generalization of the anomaly polynomial~\eqref{eq:hypvecmixg} to the higher-rank case we are considering here, and it takes the following form,
\begin{equation}\label{eq:mixedanomhrk}
\big(k_{(g^2)^2}\big)_{ab} = - u_a \delta_{ab} + H_{ab}~,\qquad  u_a > 0~, \qquad H_{ab} = H_{ba} \geq 0~.
\end{equation}
Here $u_a$ is the group-theory coefficient defined in \eqref{eq:ugdef} for the gauge algebra $\mathfrak{g}_a$, and it encodes the contribution of the $\mathfrak{g}_a$ vector multiplets to $\big(k_{(g^2)^2}\big)_{ab}$. The symmetric matrix $H_{ab}$ encodes the hyper-multiplet contributions. While it can be determined explicitly, we will only need the fact that all of its entries $H_{ab}$ are non-negative, i.e.~$H_{ab} \geq 0$ for all $a, b = 1, \ldots, r$. 

In order to invert the matrix $\big(k_{(g^2)^2}\big)_{ab}$ in \eqref{eq:mixedanomhrk}, we define the matrices
\begin{equation}\label{eq:ummatdef}
U = \text{diag}(u_1, \ldots, u_r)~, \qquad M = U^{-\half} \,  H \, U^{-\half}~.
\end{equation}
Here the matrix $U$ only depends on the gauge algebras $\mathfrak{g}_a$, while $M$ also depends on the hyper-multiplet matter content through $H$. Due to the properties of $u_a$ and $H_{ab}$ in \eqref{eq:mixedanomhrk} it follows that the matrix $M$ is symmetric, and that its matrix elements are non-negative,
\begin{equation}\label{eq:msympos}
M_{ab} = {1 \over \sqrt{u_a}} \, H_{ab} \, {1 \over \sqrt{u_b}}  = M_{ba} \geq 0~.
\end{equation}
Since $M_{ab}$ is real and symmetric it can be diagonalized, with real eigenvalues $m_a$. Note that the positivity of the matrix elements $M_{ab}$ in \eqref{eq:msympos} does not guarantee that the eigenvalues $m_a$ are positive. Using~\eqref{eq:ummatdef}, we can express the matrix $\big(k_{(g^2)^2}\big)_{ab}$ as
\begin{equation}\label{eq:kasu1mm}
k_{(g^2)^2} = - U^\half \left(\1 - M\right) U^\half~.
\end{equation}

Assume, for now, that all eigenvalues $m_a$ of $M_{ab}$ satisfy $|m_a| < 1$. (Below we will prove that this is indeed the case.) We can then invert $k_{(g^2)^2}$ by expanding $(\1 - M)^{-1}$ as an absolutely convergent power series,
\begin{equation}\label{eq:posmatrix}
\begin{split}
k^{-1}_{(g^2)^2} & = - U^{-\half}  (\1 - M)^{-1} U^{-\half}  = - U^{-\half}  \Big( \sum_{n = 0}^\infty M^n\Big) U^{-\half}~.
\end{split}
\end{equation}
Since the matrix elements of $M$ and $U$ are all non-negative (see \eqref{eq:ummatdef} and \eqref{eq:msympos}), we see that the same is true for the matrix elements of $-k^{-1}_{(g^2)^2}$. This allows us to substitute the lower bound $X_a \geq {11 h_a^\vee \over 12}$ from \eqref{eq:Xbound} into \eqref{eq:ascfthrk} to obtain the following lower bound for $\Delta a$, 
\begin{equation}\label{eq:delalbd}
\Delta a \geq -\frac{192}{7} \big(k_{(g^2)^2}\big)_{ab}^{-1} \left(11 h^\vee_a \over 12\right) \left(11 h^\vee_b \over 12\right) = \frac{192}{7}\sum_{n = 0}^\infty \Big(U^{-\half} \, M^n \, U^{-\half}\Big)_{ab} \left(11 h^\vee_a \over 12\right) \left(11 h^\vee_b \over 12\right)~.
\end{equation}

Since every term on the right-hand side of the lower bound \eqref{eq:delalbd} for $\Delta a$ are non-negative, we can obtain a weaker (but simpler, and sufficient) lower bound for $\Delta a$ by only keeping the $n = 0$ term in the sum,
\begin{equation}\label{eq:finaboundhrk}
\Delta a \geq \frac{192}{7}\Big( U^{-1}\Big)_{ab} \left(11 h^\vee_a \over 12\right) \left(11 h^\vee_b \over 12\right) = \frac{192}{7} \sum_{a = 1}^r {1 \over u_a} \left(11 h^\vee_a \over 12\right)^2~.
\end{equation}
This is an exact analogue of the lower bound \eqref{eq:finalbound} previously obtained in the rank-1 case. As was the case there, we see from~\eqref{eq:ascfthrk} that $a_\text{SCFT} > 0$ follows from the inquality
\begin{equation}\label{eq:daneclb}
\Delta a \geq {251 \over 210} n_V = {251 \over 210} \sum_{a = 1}^r \dim (\mathfrak{g}_a)~.
\end{equation}
The rank-1 proof relied on the fact that  $\frac{192}{7} {1 \over u_a} \left(11 h^\vee_a \over 12\right)^2 > {251 \over 210}  \dim (\mathfrak{g}_a)$ for every compact simple Lie algebra $\mathfrak{g}_a$. (See the discussion below \eqref{eq:finalbound}, and in particular \eqref{eq:liealgbound}.) Summing this inequality over $a = 1, \ldots, r$ and substituting into the lower bound \eqref{eq:finaboundhrk}, we conclude that the inequality \eqref{eq:daneclb} is indeed satisfied, and hence that $a_\text{SCFT} > 0$.

To complete the proof, we must go back and show that the eigenvalues $m_a$ of the matrix $M$ defined in \eqref{eq:ummatdef} satisfy
\begin{equation}\label{eq:mbd}
|m_a| < 1~~(a  = 1, \ldots, r)~,
\end{equation} 
so that we can invert $\1 - M$ using a convergent power-series expansion in $M$. To prove~\eqref{eq:mbd}, first note that all eigenvalues satisfy $m_a < 1$, because $\1 - M$ must be a positive-definite matrix. This follows from the expression $k_{(g^2)^2} = - U^\half \left(\1 - M\right) U^\half$ in  \eqref{eq:kasu1mm}, together with the facts that $k_{(g^2)^2}$ is a negative-definite matrix and $U$ is a positive diagonal matrix (see \eqref{eq:ummatdef}). 

We now make use of the fact \eqref{eq:msympos} that all entries of the matrix $M$ are non-negative, $M_{ab} \geq 0$. This implies that~$\tr(M^n) \geq 0$. Since we can express $\tr(M^n)$ in terms of the eigenvalues $m_a$ of $M$, we obtain the inequality
\begin{equation}\label{eq:trmineq}
\tr(M^n) = \sum_{a = 1}^r m_a^n = \sum_{|m_a| < 1} m_a^n + \sum_{m_a \leq -1} m_a^n \geq  0~.
\end{equation}
Here we have partitioned the sum over eigenvalues $m_a$ into those satisfying $|m_a| < 1$ and those that satisfy $m_a \leq -1$. (We have already argued that all eigenvalues must satisfy $m_a < 1$.) If we take $n \rightarrow \infty$, the sum over eigenvalues $|m_a| < 1$ vanishes and \eqref{eq:trmineq} reduces to $\sum_{m_a \leq -1} m_a^n\geq 0$. If there are any eigenvalues satisfying $m_a \leq -1$, then the left-hand side of this inequality is strictly negative for odd $n$. The only way to avoid a contradiction is to conclude that there are no such eigenvalues. Thus all eigenvalues satisfy $|m_a| < 1$, which proves \eqref{eq:mbd}. 

To summarize, we have proved that all unitary 6d SCFTs with a tensor branch have positive $a$-type Weyl anomaly, $a_\text{SCFT} > 0$.

Let us illustrate some of the ingredients that enter the proof above in an example: the rank-$r$ SCFTs constructed in \cite{Brunner:1997gk,Blum:1997mm,Brunner:1997gf}. On their tensor branches, they are described by a linear $\mathfrak{su}(n)$ quiver gauge theory associated with the $A_r$ Dynkin diagram (see for instance~\cite{Douglas:1996sw}), i.e.~the gauge algebra is $\oplus_{a = 1}^r \mathfrak{su}(n)_a$. There are bi-fundamental hyper multiplets for every link in the quiver, i.e.~for every pair of adjacent $\mathfrak{su}(n)_a$ and $\mathfrak{su}(n)_{a+1}$ gauge nodes (for $a = 1, \ldots, r-1$). In addition, there are $N$ additional fundamental flavors for the $\mathfrak{su}(n)_1$ and $\mathfrak{su}(n)_r$ nodes at the ends of the quiver. Every gauge node has $2N$ fundamental flavors, as is required for cancelling the irreducible gauge anomalies.\footnote{~These SCFTs can be embedded into LSTs~\cite{Intriligator:1997dh}. At low energies on their tensor branches, these LSTs flow to circular quiver gauge theories associated with the affine  Dynkin diagram $\widehat A_r$. There is an extra $\mathfrak{su}(n)_0$ gauge node, and the extra flavors at the ends of the $A_r$ quiver arise from bi-fundamentals between this extra $\mathfrak{su}(n)_0$ gauge node and the $\mathfrak{su}(n)_1$ or $\mathfrak{su}(n)_r$ nodes. The $\mathfrak{su}(n)_0$ gauge node is not paired with a tensor multiplet and decouples from the $A_r$ SCFT at low energies.} The Dirac pairing $\hat \Omega_{ab}$, and hence the matrix $\big(k_{(g^2)^2}\big)_{ab}$ of reducible gauge anomalies, is given by the $A_r$ Cartan matrix $C_{ab}^{(A_r)}$,
\begin{equation}
\hat \Omega_{ab} = - \big(k_{(g^2)^2}\big)_{ab} = C_{ab}^{(A_r)} = \begin{cases} 2 \qquad (a=b) \\ -1 \hskip14pt  (|a-b| = 1) \\ 0 \qquad (\text{otherwise}) \end{cases}~.
\end{equation}
Comparing with \eqref{eq:kasu1mm}, we see that $C^{(A_r)} = U^\half \left(\1- M\right) U^\half$ with $U_{ab} = 2 \delta_{ab}$, while the only non-zero entries of $M$ are $M_{a, b = a\pm 1} = \half$. The eigenvalues $m_a$ of the matrix $M$ are given by $m_a=\cos (a\pi /(r+1)) ~(a=1, \ldots, r)$ and hence satisfy the bound $|m_a|<1$ in \eqref{eq:mbd} above. Also note that the elements of the inverse Cartan matrix $(C^{(A_r)})^{-1}_{ab} = \min\{a,b\} - {ab \over r+1} > 0$ are strictly positive, in line with the discussion below~\eqref{eq:posmatrix}.

\section*{Acknowledgements}\noindent We are grateful to M.~Del Zotto, E.~D'Hoker, D.~Freed, Z.~Komargodski, G.~Moore, N.~Nekrasov, K.~Ohmori, N.~Seiberg, and S.~Shatashvili for discussions. TD is supported by a DOE Early Career Award under DE-SC0020421, as well as by a Hellman Fellowship and the Mani L. Bhaumik Presidential Chair in Theoretical Physics at UCLA. KI is supported by DOE grant DE-SC0009919, the Dan Broida Chair, and by Simons Investigator Award 568420.


\bibliographystyle{utphys}
\bibliography{references}

\end{document}